\documentclass{iopjournal}

\usepackage{bm}
\usepackage{amssymb} 
\usepackage{amsthm}
\usepackage{amsmath}
\usepackage{mathabx}
\usepackage{mathtools}
\usepackage{algorithm}
\usepackage{algpseudocode}
\usepackage[table]{xcolor} 
\usepackage{graphicx,float}
\usepackage{booktabs,multirow,array}
\usepackage{siunitx}
\usepackage{lmodern}
\usepackage{float}
\usepackage[sort&compress,numbers]{natbib}
\usepackage{hyperref}
\usepackage[dvipsnames]{xcolor}
\usepackage[table]{xcolor} 
\usepackage{array}
\usepackage{mathrsfs}
\usepackage[normalem]{ulem}
\usepackage{graphicx}
\usepackage{subcaption}
\usepackage{float} 
\usepackage{bm}
\usepackage{booktabs} 
\usepackage{multirow} 
\usepackage{array} 
\usepackage{tabularx}
\begin{document}

\title{Scalable Data-Driven Basis Selection for \\
Linear Machine Learning Interatomic Potentials}

\author{Tina Torabi$^{1,*}$, Matthias Militzer$^2$,  Michael P. Friedlander$^3$ and Christoph Ortner$^{1}$}

\affil{$^1$Department of Mathematics, University of British Columbia, Vancouver, V6T1Z2, BC, Canada}

\affil{$^2$The Centre for Metallurgical Process Engineering, University of British Columbia, Vancouver, V6T 1Z4, BC, Canada}

\affil{$^3$Department of Computer Science, University of British Columbia, Vancouver, V6T1Z2, BC, Canada}

\affil{$^*$Corresponding author}

\email{torabit@math.ubc.ca}

\begin{abstract}
Machine learning interatomic potentials (MLIPs) provide an effective approach for accurately and efficiently modeling atomic interactions, expanding the capabilities of atomistic simulations to complex systems. However, {\it a priori} feature selection leads to high complexity, which can be detrimental to both computational cost and generalization, resulting in a need for hyperparameter tuning. We demonstrate the benefits of active set algorithms for automated data-driven feature selection. The proposed methods are implemented within the Atomic Cluster Expansion (ACE) framework. Computational tests conducted on a variety of benchmark datasets indicate that sparse ACE models consistently enhance computational efficiency, generalization accuracy and interpretability over dense ACE models. An added benefit of the proposed algorithms is that they produce entire paths of models with varying cost/accuracy ratio.
\end{abstract}

\section{Introduction}

Accurate modeling of interatomic potentials is central to materials simulation. The precision with which interatomic potentials are derived directly impacts the reliability of atomic scale simulations, which are crucial for probing material behavior under diverse conditions~\cite{BernCsyani2019}.

Historically, empirical models such as the embedded-atom method (EAM)~\cite{DAW1993251} and Tersoff potentials~\cite{PhysRevB.37.6991} have provided computationally efficient means of modeling atomic interactions. However, these models often lack the accuracy and flexibility to describe complex or diverse material and molecular systems. In contrast, first-principles methods like density functional theory (DFT)~\cite{burke} offer highly accurate representations of atomic interactions but are computationally expensive, restricting their use to relatively small systems or short timescales.

Recent advances in machine learning (ML) have transformed the development of interatomic potentials by combining the accuracy of first-principles methods with the computational efficiency of empirical models. 
Machine learning interatomic potentials (MLIPs) employ rich descriptors of local atomic environments to construct systematic parameterizations of the potential energy, which is then fitted to match DFT predictions on small snapshots of the system(s) of interest. Early examples include permutation-invariant polynomials~\cite{Braams01102009}, neural networks of atom-centered symmetry functions~\cite{10.1063/1.3553717}, Gaussian approximation potentials~\cite{PhysRevLett.104.136403, Bart_k_2013}, the spectral neighbor analysis method (SNAP)~\cite{THOMPSON2015316}, moment tensor potentials (MTP)~\cite{mtp}, and the atomic cluster expansion (ACE)~\cite{PhysRevB.99.014104}. Recently, equivariant graph neural network techniques \cite{PhysRevLett.120.145301, gnns, mfdncs, graph_nns} have been gaining increasing traction due to their ability to model vastly more complex systems. However, the focus of the present work will be on linear models; specifically ACE.

For large linear MLIPs models, achieving a balance between model complexity, interpretability, and generalizability remains a challenge~\cite{Deringer2018, https://doi.org/10.1002/adma.201902765}.
Sparse selection of features can in principle lead to minimal yet informative descriptions.
Such models tend to enhance interpretability, scalability, and generalization to unseen configurations. 
Several previous studies have explored sparsification strategies for linear MLIPs. An early example of sparsification in this context appears in the MTP model~\cite{mtp}, where a matching pursuit algorithm with crossover and local search was used to minimize an objective with an $\ell_0$ regularization term, to train a potential for tungsten. While sparsity was not the primary focus, this work demonstrated that a minimal basis could be identified without sacrificing accuracy. 
Seko et al.~\cite{PhysRevB.90.024101, Seko_2015} introduced the use of the least absolute shrinkage and selection operator (LASSO)~\cite{51791361-8fe2-38d5-959f-ae8d048b490d} to automatically select basis functions from a large set of analytical radial descriptors. They demonstrated for elemental metals Na and Mg that a sparse potential energy surface can achieve high accuracy for energies and forces, while reducing the number of basis functions compared to neural network fits. Their focus was on showing that sparsity improves efficiency without sacrificing accuracy, using single-element systems and a fixed descriptor form.

Benoit et al.~\cite{Benoit_2021} studied the transferability of a linearized potential for Au–Fe nanoparticles, using LASSO regression with the Least Angle Regression scheme (LARS)~\cite{Efron_2004}, together referred to as LassoLARS. They applied this method to select descriptors from a large candidate pool, varying the penalty parameter to control model complexity. They found that adding more descriptors improved training-set accuracy but often reduced accuracy on unseen configurations, and suggested combining descriptor families to mitigate this effect. The aim of their study was to measure and interpret the complexity–transferability trade-off for a single challenging system, not to develop a general-purpose sparse MLIPs fitting methodology. In a related context, Goldman et al.~\cite{10.1063/5.0141616} employed $\ell_1$-regularized regression (LASSO/LARS) to sparsify the linear ChIMES many-body model.

Automatic relevance determination (ARD)~\cite{MacKay1996}, an empirical Bayes method that can also be viewed as a sparsification technique, was  used in the context of active learning of ACE models~\cite{Oord2022HyperactiveLF}. 

In the present work, we explore the use of two alternative sparse optimization methods, active set basis pursuit (ASP)~\cite{friedlander2012dual}, orthogonal matching pursuit (OMP)~\cite{friedlander2012dual, 342465}, and compare them against ARD and various "dense" least squares solvers. ASP and OMP generate entire "paths" of MLIP models with increasing complexity in a single run, showing how accuracy and sparsity change as more basis functions are added. These active-set solvers avoid manual hyperparameter tuning by adjusting the regularization parameter automatically at each iteration so that exactly one additional basis function is selected. This approach gives full control over model complexity, removes the need for repeated cross-validation, and provides finer resolution in the accuracy–complexity trade-off. Our focus is on demonstrating the practical utility of sparse regression for linear MLIPs, and ACE in particular, especially analyzing how model quality varies under different sparse regression strategies.

Our computational tests on three benchmark problems demonstrate several advantages. The active-set sparse solvers improve generalization, yielding lower test errors than non-sparse methods, while requiring minimal user intervention and hyperparameter tuning. Moreover, the selected features show no clear or predictable pattern, emphasizing the benefit of data-driven feature selection over heuristic or pre-defined approaches.

\section{Methods}
\subsection{Machine Learning Interatomic Potentials (MLIPs)}\label{sec:mlips}

An atomic configuration is described by a set  
\[
\mathbf{R} := \{\bm{r}_1, \ldots, \bm{r}_N\} \in \mathbb{R}^{3N},
\]
where each \(\bm{r}_j\) is the position vector of atom \(j\). Defining the displacement between atoms \(i\) and \(j\) as  
\(
\bm{r}_{ij} = \bm{r}_j - \bm{r}_i,
\)
the atomic environment around a reference atom \(i\) within a cutoff radius \(r_c\) is given by
\[
\mathbf{R}_i = \{ (\bm{r}_{ij}, Z_j) \mid j \neq i, r_{ij} \leq r_{\rm cut} \},
\]
where \(Z_j\) is the atomic number of atom \(j\), and \(r_{ij} = \|\bm{r}_{ij}\|\) is the interatomic distance. An atom is specified by $x_i = ({\bm r}_i, Z_i)$ and a bond between atoms $i$ and $j$ is specified by $x_{ij} = ({\bm r}_{ij}, Z_i, Z_j)$.

A common approach to constructing interatomic potentials is to decompose the total energy \(E\) of the system as a sum of the site energies: \(E = \sum_i \epsilon_i.
\)
In linear models, each site energy \(\epsilon_i\) is expressed as a linear combination of invariant basis functions that encode information about the local environment (e.g., interatomic distances, angles, etc.) within a cutoff radius \(r_{\text{c}}\).

The Atomic Cluster Expansion (ACE) framework provides a systematic construction of such models by expanding the site energies in terms of symmetric polynomials, organized by correlation order (or, equivalently, body order) as follows: For a given maximal correlation order \(N_{\rm max} \in \mathbb{N}\), the ACE potential is formulated as

\begin{equation}
\epsilon_i = \sum_{n=0}^{N_{\rm max}} \sum_{j_1 < j_2 < \dots < j_n} V^{(n)}(x_{ij_1}, x_{ij_2}, \dots, x_{ij_n}), \nonumber
\end{equation}
where each many-body interaction \(V^{(n)}\) is expanded in a tensor product basis
\begin{equation}
V^{(n)}(x_{ij_1}, \ldots, x_{ij_n}) = \sum_{\mathbf{k}} c_{\mathbf{k}}^{(Z_i)} \prod_{s=1}^{n} \phi_{k_s}(x_{ij_s}), \label{eq:parm}
\end{equation}
where \(\mathbf{k} = (k_1, \dots, k_p)\) and each index \(k_s\) corresponds to a tuple \( (\bm n_s, \bm \ell_s, \bm m_s) \) 
and the basis functions are defined by
\begin{equation}
\phi_{\bm n \bm \ell \bm m}(x_{ij}) = R_{\bm n \bm l}(r_{ij}, Z_i, Z_j) Y_{\bm \ell}^{\bm m}(\hat{\bm{r}}_{ij}).\nonumber
\end{equation}
Here,  \(Y_{\bm \ell}^{ \bm m}\) denote the complex or real spherical harmonics, while $R_{\bm n\bm l}$ are radial basis functions with significant freedom of choice. Detailed implementation aspects, including the choice of the radial basis \(R_{\bm n \bm l}\), can be found in~\cite{10.1063/5.0158783} and Appendix~\ref{app:ACE}.

The invariant basis can then be expressed as:
\begin{equation}
B_{{\bm n} {\bm \ell} q} (\{x_{ij}\}) = \sum_{{\bm m}} \mathcal{C}^{{\bm n} {\bm \ell} q}_{\bm m} 
\prod_{s} \sum_j \phi_{\bm m_{s} \bm \ell_{s} \bm m_{s}}(x_{ij})
\end{equation}
where \(\mathcal{C}^{\bm n \bm \ell q}_{\bm{m}}\) are generalized Clebsch-Gordan coefficients, ensuring $O(3)$-invariance of $B_{{\bm n} {\bm \ell} q}$. The index $q = 1, \dots, n_{\bm n \bm l}$ enumerates all invariant couplings; see \cite{2019-ship1} for more details. 

This leads to the final linear parameterization of the ACE potential
\begin{equation}
\epsilon_i = \sum_{B \in \mathbf{B}} c_B B(\{x_{ij}\}),\nonumber
\end{equation}
where \(c_B\) are the model parameters and \(\mathbf{B}\) is a finite set of basis functions

\begin{align}
\mathbf{B} := \Big\{ B_{{\bm n} {\bm \ell} q} \,\Big|\, & ({\bm n}, {\bm \ell}) \in \mathbb{N}^{2N} \text{ (ordered)}, \; \sum_{\alpha} \bm \ell_{\alpha} \text{ even}, \; \nonumber \; \sum_{\alpha} m_\alpha = 0, \; q = 1, \ldots, n_{{\bm n}{\bm \ell}}, \quad \nonumber \\ 
    & 
    \sum_\alpha (\bm \ell_\alpha + \bm n_\alpha) \leq D_{\rm tot}, \; N \leq N_{\text{max}} 
    \Big\}. \nonumber
\end{align}

The completeness of this representation implies that, as the approximation parameters $N_{\rm max}$ (correlation order), $r_{\rm cut}$ (cutoff radius), and $D_{\rm tot}$ (expansion precision) tend to infinity, the model is capable of representing any arbitrary potential~\cite{2019-ship1}. 
Although these parameters provide a systematic way to refine the approximation, manually selecting them through heuristics or grid searches introduces a major bottleneck in model development. Even if optimal parameters can be selected, this does not result in an optimal or even near-optimal choice of basis, which would require selection of individual basis functions.
Rather than fine-tuning hyperparameters manually, the goal of our framework is to eliminate the need for exhaustive hyperparameter tuning by performing automatic feature selection, ensuring that only the most informative basis functions are directly selected during parameter estimation. We will demonstrate in several examples that this data-driven approach increases automation, reduces computational overhead and enhances generalization, stability, and interpretability of the resulting interatomic potential.

\subsubsection*{Parameter Estimation and Regularization}
To estimate the coefficients \(\bm{c}\), a training dataset of atomic configurations \(\mathfrak{R}=\{\mathbf{R}\}\) is required. For each configuration \(\mathbf{R}\), target potential energy \(\mathcal{E}_\mathbf{R}\),  forces \(\mathcal{F}_\mathbf{R}\) and virials \(\mathcal{V}_\mathbf{R}\) are provided, usually from DFT calculations. The model parameters are then obtained by minimizing a least-squares loss function
%

\begin{align*}
    L(\bm{c}) &= \sum_{\mathbf{R} \in \mathfrak{R}} \Bigl( 
    w_{E, \mathbf{R}}^2 \left| E(\bm{c}; \mathbf{R}) - \mathcal{E}_\mathbf{R} \right|^2 + \; w_{F, \mathbf{R}}^2 \left| F(\bm{c}; \mathbf{R}) - \mathcal{F}_\mathbf{R}\right|^2 + \; w_{V, \mathbf{R}}^2 \left| V(\bm{c}; \mathbf{R}) - \mathcal{V}_\mathbf{R} \right|^2 
    \Bigr). \nonumber
\end{align*}

Here, the weights \(w_{E, \mathbf{R}}\), \(w_{F, \mathbf{R}}\), and \(w_{V, \mathbf{R}}\) assign varying levels of importance to different observations and structures. In all our tests we use the defaults published in \cite{10.1063/5.0158783}. The coefficients $\bm{c}$ are obtained by solving the minimization problem:
\begin{equation}
    \bm{c} = \arg\min_{\bm{c}} \| \mathbf{W}(\bm{y} - \mathbf{A}\bm{c}) \|_2^2, \nonumber
\end{equation}
where \(\bm{y}\) collects the observed data (energies, forces, stresses), \(\mathbf{A}\) is the design matrix composed of basis function values, and \(\mathbf{W}\) is a diagonal matrix that can be used to specify chosen weights.

To avoid overfitting, one typically employs regularized modifications such as ridge or Bayesian regression. In the context of ACE MLIPs, a natural approach to regularize is to encourage ``smoothness'' of the fitted potential, or equivalently, avoid oscillatory behavior that are common when overfitting polynomials. Concretely, we employ an \emph{algebraic smoothness prior} introduced in \cite{2021-apxsym, 10.1063/5.0158783}. Each basis function, indexed by \(\{n_t, l_t\}\), is assigned a smoothness cost:
\begin{equation}
\gamma_{{\bm n} {\bm \ell}{\bm m}}(p) = \sum_t \bigl(n_t^p + w_L\,l_t^p\bigr),\nonumber
\end{equation}
where $w_L$ balances the contributions of the radial and angular components; we again use defaults from \cite{10.1063/5.0158783}. These costs are assembled into a diagonal matrix \(\Gamma\) (with \(\Gamma_{kk} = \gamma_k\)) and incorporated into the regularized minimization:
\begin{equation}  \label{eq:reg}
\arg \min_{\bm{c}} \left\| \mathbf{W} (\bm{y} - \mathbf{A} \bm{c}) \right\|^2
+ \lambda \left\|\Gamma \bm{c} \right\|^2.\nonumber 
\end{equation}
Higher values of \(\lambda\) or \(\gamma_k\) favor smoother, less oscillatory potentials. The \texttt{ACEpotentials.jl} package~\cite{10.1063/5.0158783} provides additional smoothness priors—such as the exponential and Gaussian priors—that offer varying degrees of penalization on oscillatory terms.

The  \texttt{ACEpotentials.jl} package~\cite{10.1063/5.0158783} also offers a range of solvers tailored for different dataset sizes and computational requirements. Once the model parameters are determined, the next step is to validate the potential on a test dataset. Common metrics for evaluating accuracy include the root mean square error (RMSE) and mean absolute error (MAE) for energies, forces, and stresses. Following successful validation, the model can be exported in various formats for subsequent use in molecular dynamics simulations.

\subsection{Sparse recovery methods}\label{sec:sparsesolver}

Least-squares (LS) problems, such as \eqref{eq:reg}, arise in numerous applications across signal processing, data science, and machine learning~\cite{JMLR:v18:15-586, doi:10.1137/S0895479896298130}. In a standard LS problem, the objective is to solve for a vector \( x \in \mathbb{R}^n \) that minimizes the residual \( \|Ax - b\|_2^2 \), where \( A \in \mathbb{R}^{m \times n} \) is a known matrix, often termed a dictionary, and \( b \in \mathbb{R}^m \) is the target vector. Sparse solutions, where most entries in \( x \) are zero or near-zero, are particularly desirable in scenarios where only a small subset of predictors or signal components is relevant. Sparse solutions have important implications in model selection as they enable the identification of minimal predictor subsets that improve both interpretability and model generalization~\cite{Tibshirani1996RegressionSA}. In signal processing, they facilitate signal reconstruction from limited measurements by assuming the signal is representable in a sparse basis~\cite{mallat2009wavelet}. Compressed sensing leverages sparsity to efficiently acquire and reconstruct signals from a reduced set of samples, thereby reducing data acquisition costs~\cite{candes2006compressive}. 

In optimization terms, the goal is to identify the smallest possible subset of \emph{features} from the set represented by the columns of the matrix $A$. A sparse solution \( x \) thus corresponds to selecting the minimal set of features that best approximates the target vector \( b \). The Sparse Approximate Solution (SAS) problem can be formulated as follows: Given a matrix \( A \in \mathbb{R}^{m \times n} \), where \( A \) has \( m \) rows and \( n \) columns, a target vector \( b \in \mathbb{R}^m \), and a tolerance \( \epsilon > 0 \), find a vector \( x \) that satisfies
\[
\|Ax - b\|_2 \leq \epsilon
\]
and has the fewest possible nonzero entries among all vectors satisfying this condition. The SAS problem is known to be NP-hard~\cite{doi:10.1137/S0097539792240406}. However, practical alternatives exist, often classified into three main categories: convex relaxations, non-convex optimization techniques, and greedy algorithms. In what follows we will briefly review a few of these alternatives.

\subsubsection{Convex relaxations} \label{sec:cvxs}

Convex relaxations offer tractable alternatives to SAS and, under certain conditions, can approximate or even recover the exact sparsest solution \cite{https://doi.org/10.1002/cpa.20124}. Standard formulations include Basis Pursuit (BP)~\cite{doi:10.1137/S1064827596304010}, Basis Pursuit Denoising (BPDN)~\cite{doi:10.1137/S003614450037906X}, and LASSO~\cite{51791361-8fe2-38d5-959f-ae8d048b490d}, whose duals reduce to linear or quadratic programs~\cite{doi:10.1137/080714488}.
Consider the following primal and dual optimization problems:
\begin{equation}\label{eq:primal-dual}
\begin{aligned}
\textbf{Primal:} \quad & \underset{x, y}{\text{minimize}} \quad \|x\|_1 + \frac{1}{2}\lambda \|y\|_2^2 \\
& \text{subject to} \quad Ax + \lambda y = b, \\[1em]
\textbf{Dual:} \quad & \underset{y}{\text{maximize}} \quad b^T y - \frac{1}{2}\lambda \|y\|_2^2 \\
& \text{subject to} \quad -\mathbf{1} \leq A^T y \leq \mathbf{1},
\end{aligned}
\end{equation}

where $\lambda \geq 0$ is the regularization parameter. Setting $\lambda=0$ yields BP, while $\lambda>0$ gives BPDN. Optimal pairs $(x,y)$ satisfy the KKT conditions~\cite{KuhnTucker+1951+481+492, karush39minima}.

Algorithms for these problems include the BP-simplex method~\cite{doi:10.1137/S003614450037906X,lp2002}, which iteratively swaps basis elements, and the Homotopy method~\cite{10.1093/imanum/20.3.389}, which solves a sequence of LASSO problems with decreasing regularization parameters $\tau_k$:

\[
\underset{x}{\text{minimize}}\;  \|Ax - b\|_2^2 \quad \text{subject to} \quad \|x\|_1 \leq \tau_k,
\] 

 updating the solution at breakpoints where the active set changes (see Appendix~\ref{sec:homotopy}).

The orthogonal matching pursuit (OMP)~\cite{342465} algorithm is a greedy algorithm that iteratively builds a solution to the problem
\[
\underset{x}{\text{minimize}} \; \|A x - b\|_2^2 \quad \text{subject to} \quad \|x\|_0 \leq k
\]
by selecting the column of A most correlated with the current residual. After selecting a column, OMP updates the solution by solving a least-squares problem using the selected columns and computes a new residual orthogonal to the selected columns. OMP continues this process, adding one column at a time, until a stopping criterion is met, such as achieving a desired residual norm or reaching a predetermined number of iterations. The details of OMP will be discussed in Appendix~\ref{sec:omp}. 
Friedlander and Saunders~\cite{friedlander2012dual} introduced the parameterization~\eqref{eq:primal-dual} of the BPDN problem, based on a dual active-set approach suitable for large-scale problems, as it requires only matrix-vector products. Their algorithm iteratively identifies the active set of constraints, solves a least-squares subproblem on those constraints, and updates the solution estimate by moving along a descent direction. The algorithm continues until it reaches a stationary point where the objective gradient is a linear combination of the gradients of the active constraints, indicating optimality. We describe this algorithm in more detail in Appendix~\ref{sec:bpdual}.

\textbf{Remark I.}
In the context of fitting MLIPs, the central goal is to identify the most influential basis functions. The path of solutions as the regularization parameter varies is crucial because it reveals the order in which basis functions enter the model and how their coefficients evolve. By tracing the full solution path, we can select a model that achieves predictive accuracy while retaining only the minimal set of meaningful basis functions.

We found that only a limited number of software packages reproduce the full solution path. In the Julia language, the main available implementations are \texttt{Lasso.jl} and \texttt{LARS.jl}, alongside our implementation of the ASP algorithm, originally developed in MATLAB and now available in Julia as \texttt{ActiveSetPursuit.jl}. Our numerical experiments revealed that \texttt{LARS.jl} and \texttt{Lasso.jl} lack robustness, efficiency, and practical usability, and they frequently fail to produce reliable results. This is demonstrated in Section~\ref{sec:comp}.

\subsubsection{Empirical Bayes methods}

Bayesian ridge regression (BRR)~\cite{Deisenroth_Faisal_Ong_2020,MacKay1992BayesianI} is a probabilistic approach to linear regression that uses a Bayesian framework to estimate the posterior distribution of the regression coefficients. By modeling the noise and prior distributions as Gaussians, BRR obtains a closed-form solution for the posterior. Automatic relevance determination (ARD)~\cite{MacKay1996} extends this framework by assigning individual precision hyperparameters to each coefficient, thereby facilitating automatic feature selection.

Given the linear model
\[
  b = Ax + \boldsymbol{\epsilon}, \nonumber
\]
where the error \( \boldsymbol{\epsilon} \sim \mathcal{N}(\mathbf{0}, \lambda^{-1} \mathbf{I}),\nonumber \)
the likelihood function is
\begin{equation}
  p(b \mid A, x, \lambda) = \left(\frac{\lambda}{2\pi}\right)^{m/2} \exp\!\left(-\frac{\lambda}{2}\|b - Ax\|^2\right).\nonumber
\end{equation}

A Gaussian prior is placed on the regression coefficients:
\begin{equation}
  p(x \mid \alpha) = \mathcal{N}(x \mid \mathbf{0}, \alpha^{-1} \mathbf{I}), \label{eq:brr}
\end{equation}
resulting in a Gaussian posterior
\begin{equation}
  p(x \mid b, A, \alpha, \lambda) = \mathcal{N}(x \mid \hat{x}, \boldsymbol{\Sigma}),\nonumber
\end{equation}
with $\boldsymbol{\Sigma} = (\alpha \mathbf{I} + \lambda A^\top A)^{-1},$ and $
  \hat{x} = \lambda \boldsymbol{\Sigma} A^\top b.$ Hyperparameters $\alpha$ and $\lambda$ are typically optimized via evidence maximization.

Automatic relevance determination (ARD) generalizes~\eqref{eq:brr} by assigning an independent precision $\alpha_i$ to each coefficient:
\begin{equation}
  p(x \mid \boldsymbol{\alpha}) = \mathcal{N}(x \mid \mathbf{0}, \mathbf{\Gamma}^{-1}),\nonumber
\end{equation}
where the covariance matrix is
\begin{equation}
  \mathbf{\Gamma} = \mathrm{diag}(\alpha_1, \alpha_2, \dots, \alpha_{N_{\text{basis}}}). \nonumber
\end{equation}
This selective regularization facilitates feature selection by effectively shrinking irrelevant coefficients toward zero.

In the context of ACE MLIPs, ARD has been used within the Hyperactive Learning (HAL) framework~\cite{Oord2022HyperactiveLF}. After fitting the model, posterior samples of the model parameters are drawn, and a committee of ACE models is formed to efficiently approximate the uncertainty in energy and force predictions. This uncertainty is then used in the HAL loop to identify configurations where the model is least certain.
For further details on the derivations and optimization, refer to Appendix~\ref{app:brr}.

\textbf{Remark II.} Sparse regression, as considered in this work, refers to selecting a small subset of basis functions from a large candidate pool to build linear models. It should not be confused with sparse Gaussian process regression~\cite{leibfried2022tutorialsparsegaussianprocesses}, where sparsity refers to data sparsity, achieved by using a limited set of inducing points to approximate the kernel and make training feasible on large datasets. It should also not be confused with tensor sketching~\cite{2022-randomembedding} which is occasionally also referred to as a form of sparsification but is an entirely different form of dimension reduction.

\section{Results}\label{sec:results}

We present a number of benchmark results to test the proposed parameter estimation methods. 
In this section, we refer to the homotopy LASSO solver and the Orthogonal Matching Pursuit solver, introduced in Section~\ref{sec:cvxs}, respectively, as  {ASP} and  {OMP}. We employ a Julia implementation {\tt ActiveSetPursuit.jl}~\cite{ActiveSetPursuitJL} of the BPdual, Homotopy, and OMP algorithms, based on~\cite{ASP}.

An artifact of the ASP(BPDN) model is its tendency to shrink coefficients toward zero as a result of the $\ell_1$-regularization. To de-bias~\cite{10.1111/rssb.12026, 10.1214/14-AOS1221} the model, we applied a post-processing step using truncated singular value decomposition (TSVD) to refine the final coefficients. TSVD mitigates this issue by projecting the solution onto a lower-dimensional subspace, filtering out small singular values that contribute to numerical instability while preserving the dominant structures in the solution. This post-processing step thus improves both the accuracy and robustness of the final coefficient estimates. Further details on the TSVD procedure and its effects on the results can be found in Appendix~\ref{app:tsvd}. Selecting the truncation parameter for the TSVD post-processing requires a validation set. The dataset splitting procedure is described in the following subsections.

Apart from test errors, a key indicator of a machine-learned interatomic potential’s quality is its stability—the ability to sustain long molecular dynamics (MD) simulations while ensuring the system remains in physically meaningful states without divergence or instability. To assess the stability of the ACE potentials, we conducted NVT MD simulations on test set configurations at 300K and 500K, using a 5-femtosecond timestep over a 1-nanosecond trajectory for metals and semi-metals, and a 1-femtosecond timestep over a 1-nanosecond trajectory for water. Our results confirmed the stability of {\em all} fitted ACE potentials under these conditions. For more details please refer to \textit{Supplementary Information}.
\begin{table}[!t]
\centering
\caption{Summary of solvers used for ACE model construction.}
\begin{tabular}{p{2cm} p{11cm}}
\toprule
\textbf{Solver} & \textbf{Description} \\
\midrule
\textbf{RRQR} & Rank‐revealing QR decomposition with random matrix sketching; removes sensitive subspaces based on the tolerance parameter $r_\mathrm{tol}$, which is related to $\lambda$ (Eq.~\ref{eq:reg})~\cite{CHAN198767}. More performant than standard QR for large datasets. \\[3pt]

\textbf{BLR} & Bayesian linear regression that selects $\lambda$ via Bayesian evidence maximization~\cite{Deisenroth_Faisal_Ong_2020, MacKay1992BayesianI}. More robust than RRQR but computationally intensive; best suited for small datasets and uncertainty quantification. \\[3pt]

\textbf{ARD} & Empirical Bayes method with hierarchical sparsity priors to prune basis functions automatically~\cite{MacKay1996}. \\[3pt]

\textbf{ASP} & Homotopy solver for the LASSO problem; refines the active basis set along a continuous path~\cite{friedlander2012dual}. Scalable and well‐suited for large datasets requiring structured sparsity. \\[3pt]

\textbf{OMP} & Greedy sparse regression algorithm that selects the most relevant basis functions iteratively~\cite{friedlander2012dual}. More computationally efficient than ASP. \\
\bottomrule
\end{tabular}
\label{tab:solvers}
\end{table}

A summary of all the solvers used within this section and their most appropriate use cases is shown in Table~\ref{tab:solvers}. This table provides a concise overview of each solver's methodology, along with guidance on when it is most applicable for different problem settings. For each set of results we compare sparse solvers to a reference dense solver employed in previous tests in \cite{10.1063/5.0158783}.

\subsection{Limited diversity material datasets}\label{sec:limit_metal}
We evaluate the performance of the ASP and OMP solvers with default parameters on several low-diversity datasets, which arise for example in automated workflows that require special purpose potentials for a limited range of tasks, in particular (hyper-)active learning. We employ the single-element benchmarks~\cite{Zuo2020} including six datasets for, respectively, Li, Mo, Ni, Cu, Si, and Ge. These elements cover a range of chemical properties (main group metals, transition metals, and semiconductors), crystal structures (bcc, fcc, and diamond), and bonding types (metallic and covalent). Each dataset comprises the element’s ground-state crystal structure, strained configurations with strains ranging from $-10\%$ to $10\%$, slab structures with Miller indices up to three, and NVT \textit{ab initio} molecular dynamics simulations of bulk supercells, both with and without a single vacancy. We refer to \cite{Zuo2020} for further details. While the datasets provide a large number of training structures, they exhibit limited diversity compared to more recent MLIPs training sets. 


For this benchmark, the training and test sets are predefined. To construct the validation set, we randomly shuffle the training data and split it into 85\% training and 15\% validation subsets.
Our starting basis size was 5292 (corresponding to a $D_{\rm tot}$ of 25 and a body order of 3). We then allowed OMP, ASP, and the ARD solver from scikit-learn~\cite{scikit-learn} to select the most relevant basis functions. We terminated  {ASP} and  {OMP} when the number of active elements hit 300 for the small model and 1000 for the large model. We also obtained the  {ARD} solutions with desired sparsity by modifying $\boldsymbol{\alpha}$. The resulting MAE errors of the potentials trained are compared to the previous results reported in~\cite{10.1063/5.0158783}.
As can be seen from the Tables~\ref{tab:Zuo_energies},~\ref{tab:Zuo_forces}, apart from Li where the RRQR solver outperforms OMP and ASP, which suggests that fine-tuning of the model parameters could be beneficial, the OMP and ASP solvers outperform all other elements' best checkpoint reported in~\cite{10.1063/5.0158783}.
\begin{table*}[!t]
\centering
\caption{Mean absolute test errors in predicted energies (meV) and forces (eV/\AA) for small ACE models ($\approx 300$ basis functions) and large ACE models ($\approx 1000$ basis functions) using different solvers, compared with the two best-performing MLIPs reported in~\cite{Zuo2020}.}
\label{tab:subtables}

\begin{subtable}{\textwidth}
\centering
\caption{Mean absolute test errors in predicted energies (meV).}
\label{tab:Zuo_energies}
\begin{tabular}{@{}lcccccccccc@{}}
\toprule
\textbf{Element} & \multicolumn{4}{c}{\textbf{Small model [meV]}} & \multicolumn{4}{c}{\textbf{Large model [meV]}} & \textbf{MTP} & \textbf{GAP} \\ 
\cmidrule(lr){2-5} \cmidrule(lr){6-9}
& RRQR & ASP & OMP & ARD & RRQR & ASP & OMP & ARD &  &  \\
\midrule
Ni & 0.416 & \textbf{0.220} & 0.232 & 0.561 & 0.340 & 0.253 & \textbf{0.241} & 0.321 & 0.42 & 0.48 \\
Cu & 0.292 & \textbf{0.232} & 0.269 & 0.826 & 0.228 & 0.217 & \textbf{0.194} & 0.315 & 0.46 & 0.41 \\
Li & \textbf{0.231} & 0.373 & 0.345 & 0.438 & \textbf{0.165} & 0.355 & 0.354 & 0.319 & 0.49 & 0.49 \\
Mo & 2.597 & 2.261 & \textbf{2.123} & 3.945 & 2.911 & 2.138 & \textbf{1.933} & 2.381 & 2.24 & 2.83 \\
Si & 3.501 & 2.444 & \textbf{2.296} & 3.891 & 2.388 & 1.981 & \textbf{1.859} & 2.344 & 2.91 & 2.21 \\
Ge & 2.594 & 2.212 & \textbf{1.849} & 2.630 & 2.162 & 2.018 & \textbf{1.936} & 2.093 & 2.06 & \textbf{1.79} \\
\bottomrule
\end{tabular}
\end{subtable}

\vspace{0.6cm}

\begin{subtable}{\textwidth}
\centering
\caption{Mean absolute test errors in predicted forces (eV/\AA).}
\label{tab:Zuo_forces}
\begin{tabular}{@{}lcccccccccc@{}}
\toprule
\textbf{Element} & \multicolumn{4}{c}{\textbf{Small model [eV/\AA]}} & \multicolumn{4}{c}{\textbf{Large model [eV/\AA]}} & \textbf{MTP} & \textbf{GAP} \\ 
\cmidrule(lr){2-5} \cmidrule(lr){6-9}
& RRQR & ASP & OMP & ARD & RRQR & ASP & OMP & ARD &  &  \\
\midrule
Ni & 0.018 & \textbf{0.014} & \textbf{0.014} & 0.026 & 0.015 & 0.014 & \textbf{0.013} & 0.019 & 0.02 & 0.01 \\
Cu & 0.007 & 0.006 & \textbf{0.005} & 0.019 & \textbf{0.005} & \textbf{0.005} & \textbf{0.005} & 0.014 & 0.01 & 0.01 \\
Li & \textbf{0.006} & 0.008 & 0.008 & 0.056 & \textbf{0.005} & 0.007 & 0.007 & 0.008 & 0.01 & 0.01 \\
Mo & 0.123 & 0.104 & \textbf{0.098} & 0.138 & 0.097 & 0.087 & \textbf{0.086} & 0.092 & 0.09 & 0.09 \\
Si & 0.086 & 0.074 & \textbf{0.072} & 0.086 & 0.066 & 0.061 & \textbf{0.059} & 0.060 & 0.07 & 0.06 \\
Ge & 0.064 & 0.059 & \textbf{0.056} & 0.072 & 0.051 & 0.049 & \textbf{0.047} & 0.049 & 0.05 & 0.05 \\
\bottomrule
\end{tabular}
\end{subtable}
\end{table*}

\begin{figure*}[htb!]
     \centering
     \begin{subfigure}[b]{0.32\linewidth} %
         \centering
\includegraphics[width=\textwidth]{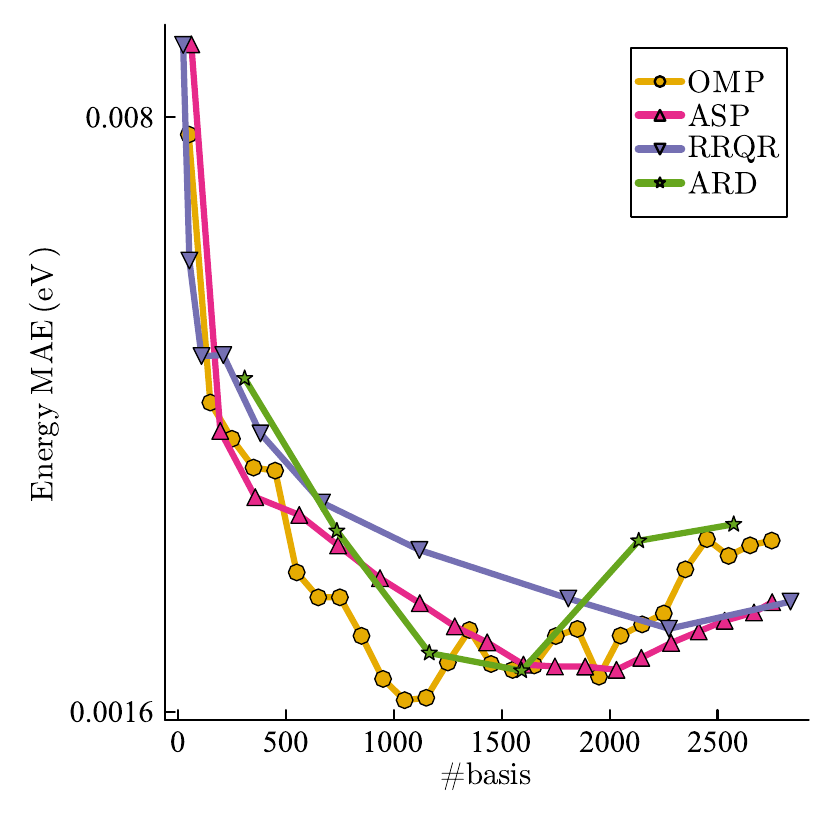}
         \caption{Mo}
         \label{fig:molr}
     \end{subfigure}
     \vspace{0.2cm} 
     \begin{subfigure}[b]{0.32\linewidth}
         \centering
\includegraphics[width=\textwidth]{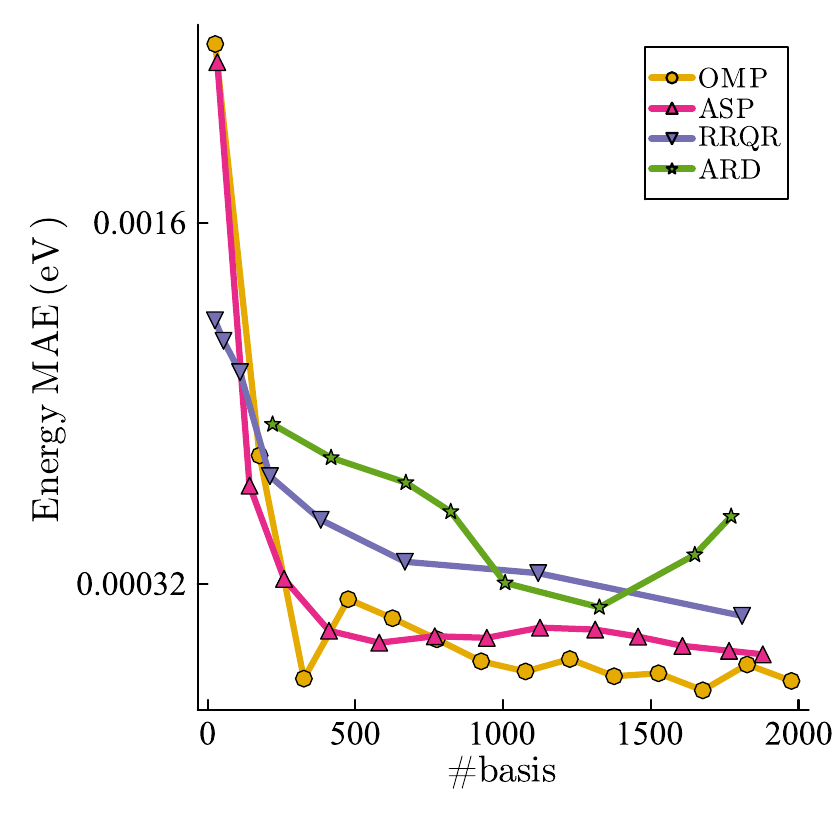}
         \caption{Ni}
         \label{fig:nilr}
     \end{subfigure}
     \vspace{0.2cm} 
     \begin{subfigure}[b]{0.32\linewidth}
         \centering
\includegraphics[width=\textwidth]{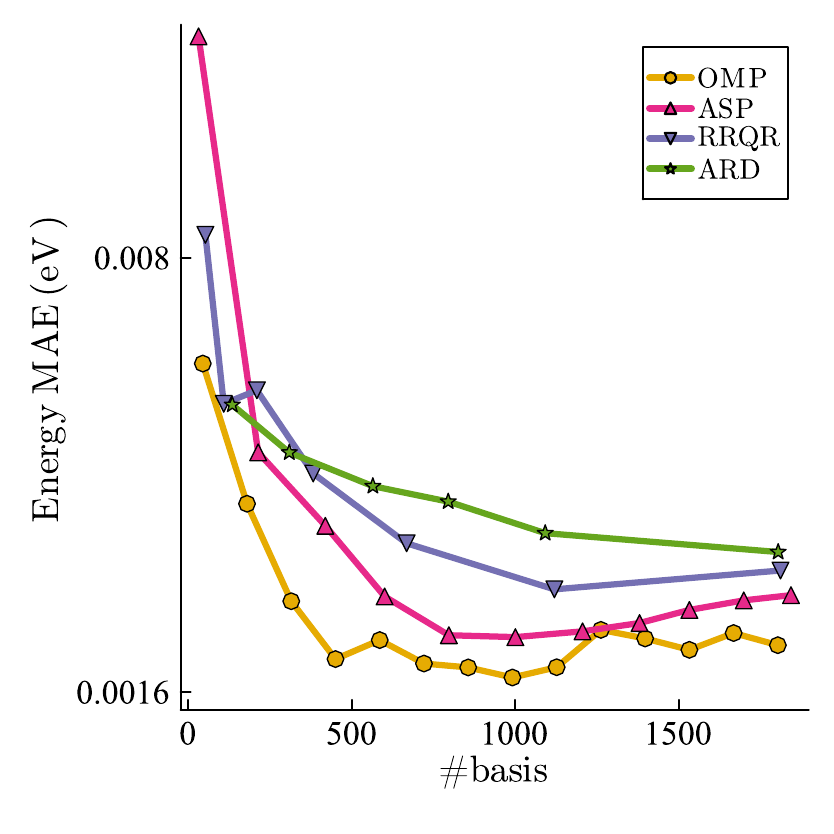}
         \caption{Si}
         \label{fig:silr}
     \end{subfigure}
    \caption{Energy MAE vs basis size for selected limited diversity datasets (cf.~\ref{sec:limit_metal}), comparing three  sparse least squares solvers (ARD, ASP, OMP) with a direct regularized least squares approach (RRQR). 
    }
    \label{fig:zuo_LR}
\end{figure*}

Figure~\ref{fig:zuo_LR} shows a comparison of the MAE for energy predictions between OMP, ASP, RRQR, and ARD for the Mo, Ni, and Si datasets as the number of active basis functions increases. 
OMP and ASP solvers consistently achieve lower MAE than  {RRQR} and  {ARD} as the basis size increases, outperforming them for the majority of the range. Among the solvers in  {ActiveSetPursuit}, while  {OMP} achieves lower errors,  {ASP} appears to be more stable, with  {OMP} exhibiting some fluctuations. {Since  {OMP} and  {ASP} significantly outperform  {ARD}, we will present results only for  {OMP} and  {ASP} in subsequent sections.}

\begin{figure*}[htb]
     \centering
     \begin{subfigure}[b]{0.32\linewidth}
         \centering
\includegraphics[width=\textwidth]{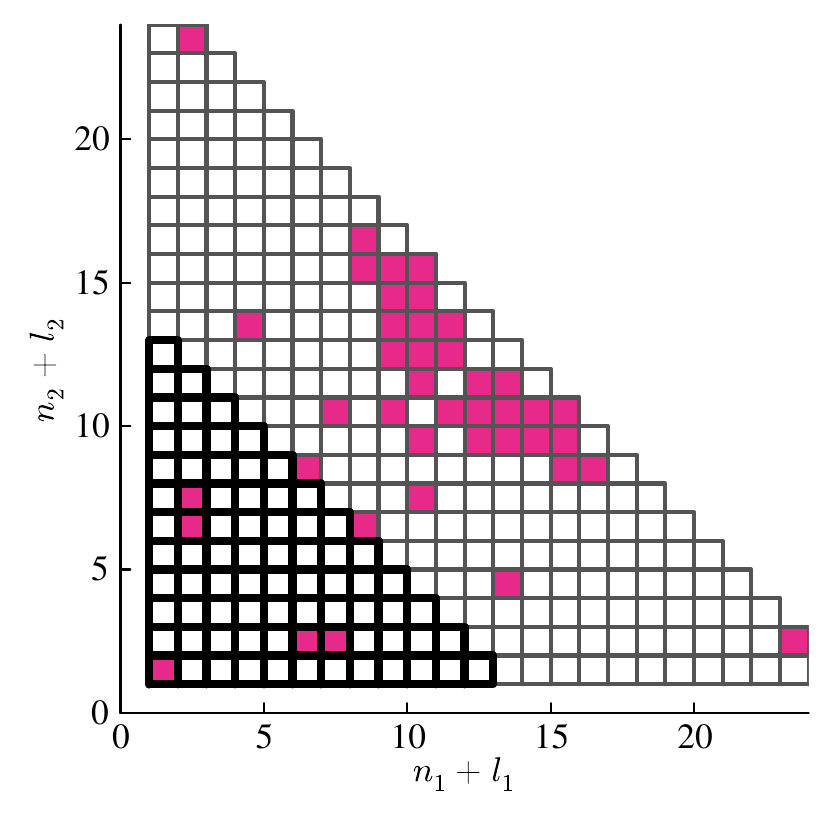}
         \caption{\#active = 300}
         \label{fig:mo2d300}
     \end{subfigure}
     \vspace{0.2cm} 
     \begin{subfigure}[b]{0.32\linewidth}
         \centering
         \includegraphics[width=\textwidth]{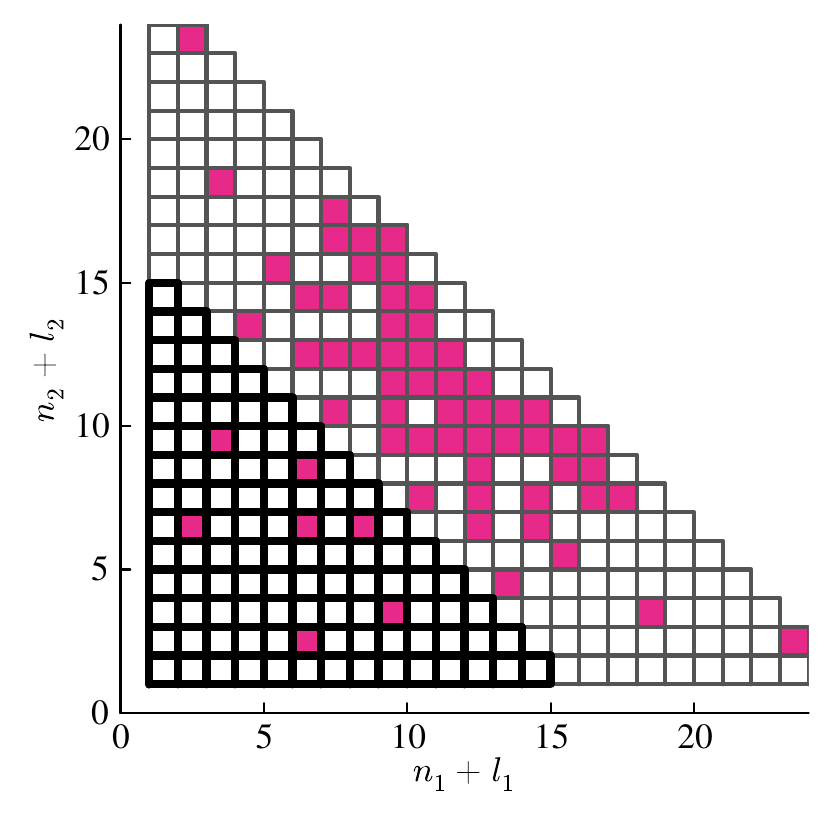}
         \caption{\#active = 500}
         \label{fig:mo2d500}
     \end{subfigure}
     \vspace{0.2cm} 
     \begin{subfigure}[b]{0.32\linewidth}
         \centering
         \includegraphics[width=\textwidth]{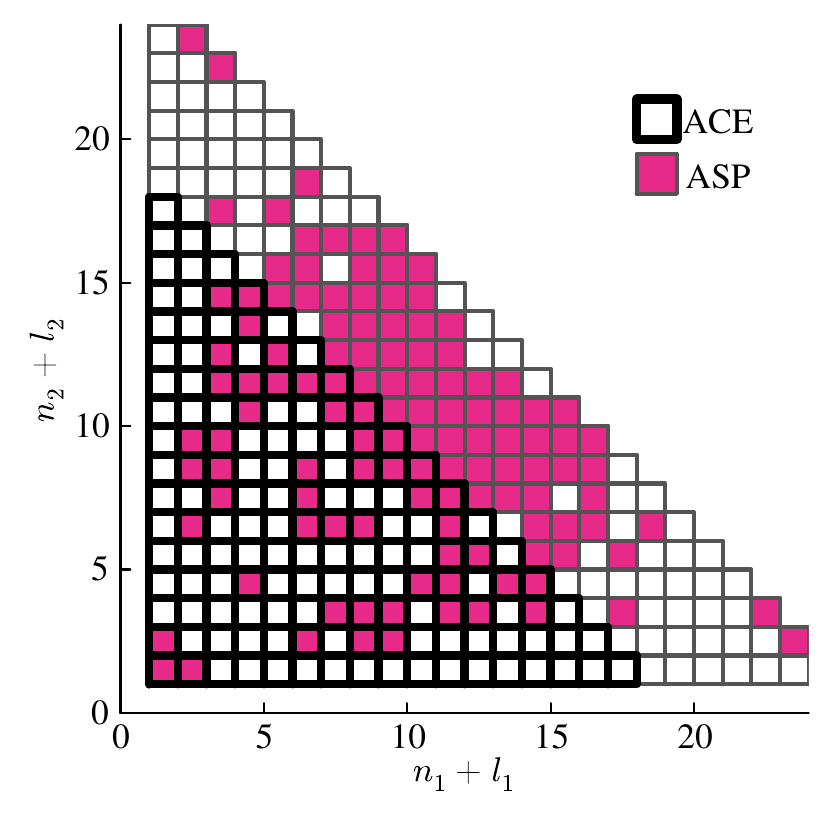}
         \caption{\#active = 1000}
         \label{fig:mo2d1000}
     \end{subfigure}
     
    \caption{Visualization of the basis functions selected for two-body interactions in the Mo dataset (cf. ~\ref{sec:limit_metal}). The figure illustrates the gradual selection process of ASP and non-sparse ACE solvers as the number of active basis functions increases. The ASP-selected basis functions (pink) show a distinct, data-driven selection pattern compared to the full ACE basis (black), demonstrating the benefits of data-driven basis selection.}
    \label{fig:mo2d}
\end{figure*}

\begin{figure*}[htb!]
     \centering
          \begin{subfigure}[b]{0.31\textwidth}
         \centering
         \qquad \qquad \qquad
         \includegraphics[width=\textwidth]{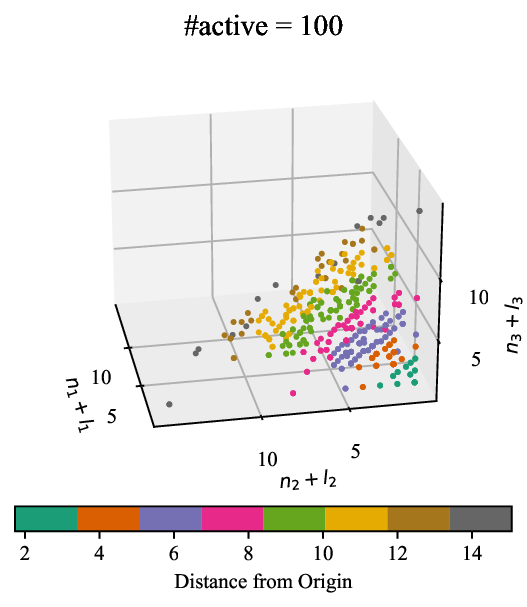}
         \label{fig:mo3d100}
     \end{subfigure}
     \begin{subfigure}[b]{0.31\textwidth}
         \centering
        \includegraphics[width=\textwidth]{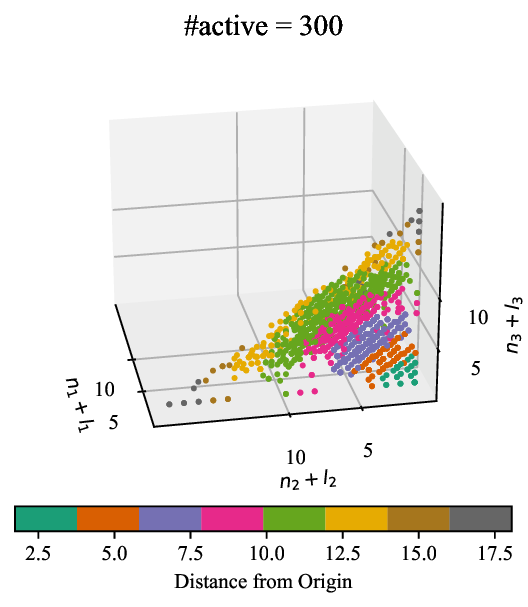}
         \label{fig:mo3d500}
     \end{subfigure}
     \begin{subfigure}[b]{0.31\textwidth}
         \centering
         \qquad \qquad \qquad
         \includegraphics[width=\textwidth]{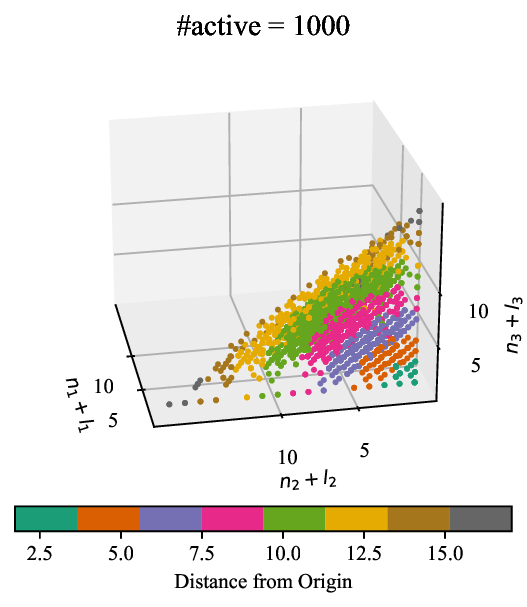}
         \label{fig:mo3d1000}
     \end{subfigure}
     
    \caption{
    Visualization of the basis functions selected for three-body interactions in the Mo dataset (cf. ~\ref{sec:limit_metal}). The figure illustrates the gradual selection process of ASP and non-sparse ACE solvers as the number of active basis functions increases. The selection is visualized in 3D, with colors indicating the distance from the origin. Similarly as with 2-correlation, ASP selects features without any a priori predictable pattern.}
    \label{fig:mo3d}
\end{figure*}

Figures~\ref{fig:mo2d} and \ref{fig:mo3d} visualize the basis functions selected by the  {ASP} solver for, respectively, the two-body and three-body interactions. In Figure~\ref{fig:mo2d}, the black squares correspond to basis functions selected by any non-sparse solver at a given basis size, while the pink squares indicate the basis functions chosen by the ASP solver.
In Figure~\ref{fig:mo3d}, the dots correspond to the basis functions chosen by ASP, separated according to their distance from the origin for visualization purposes.

The ASP solver tends to select a larger proportion of three-body basis functions compared to two-body ones. The patterns of selected basis functions are non-intuitive and deviate significantly from any common {\it a priori} basis selection mechanism such as total degree or hyperbolic cross selections. A speculative explanation is that two-body features can be approximately represented in terms of three-body features, hence the sparse basis selection reduces the resulting ill-conditioning. Regardless of the explanation for the selected basis, this result highlights the significant potential benefits of applying data-driven basis selection methods. 

We performed an additional test on the limited-diversity elements dataset, starting the OMP solver with varying initial basis sizes to examine their effect on model accuracy when the solvers were terminated at approximately 1000 basis functions. The results are shown in Table~\ref{tab:initial_basis_zuo}. In some cases, we observe a slight decrease in energy error with increasing initial basis size, while the force error remains stable. However, these changes are small, suggesting that the starting basis size does not have a significant impact on accuracy. This indicates that users can start the solvers with the largest basis size suitable for their application. In our own tests we generally use a starting basis approximately twice as large as the target basis size.

\begin{table*}[!t]
\centering
\small
\setlength{\tabcolsep}{8pt}
\renewcommand{\arraystretch}{1.2}
\caption{Energy and force mean absolute errors (MAE) for different numbers of starting basis functions for the Mo, Ni, and Si datasets. All models select 1000 features. Increasing the starting basis size gives consistent moderate improvements in fit accuracy, suggesting that it should be chosen primarily based on affordable computational cost for parameter estimation.}
\label{tab:initial_basis_zuo}
\begin{tabular}{@{}lccc@{}}
\toprule
\textbf{Element} & \textbf{\# Starting basis} & \textbf{Energy MAE (meV/atom)} & \textbf{Force MAE (meV/\AA)} \\
\midrule
\multirow{5}{*}{Mo}
& 1429 & 2.756 & 0.092 \\
& 2275 & 1.952 & 0.091 \\
& 3515 & 2.101 & 0.090 \\
& 5292 & 1.933 & 0.089 \\
& 7777 & 2.242 & 0.088 \\[3pt]
\midrule
\multirow{5}{*}{Si}
& 1429 & 2.637 & 0.065 \\
& 2275 & 2.403 & 0.063 \\
& 3515 & 1.775 & 0.061 \\
& 5292 & 1.859 & 0.062 \\
& 7777 & 1.558 & 0.062 \\[3pt]
\midrule
\multirow{5}{*}{Ni}
& 1429 & 0.297 & 0.014 \\
& 2275 & 0.217 & 0.014 \\
& 3515 & 0.229 & 0.014 \\
& 5292 & 0.241 & 0.013 \\
& 7777 & 0.276 & 0.013 \\
\bottomrule
\end{tabular}
\end{table*}

\subsubsection{Comparison with Other Path-Tracing Solvers}\label{sec:comp}

To explain our choice of the ASP and OMP solvers, we compared them against two available software packages that generate LASSO paths. The results are summarized in Table~\ref{tab:software_Zuo}. As shown, the MAEs obtained from potentials fitted with \texttt{LARS.jl}~\cite{KornblithJohnson2025LARSjl} and \texttt{Lasso.jl}~\cite{JuliaStatsLasso} are significantly higher than those achieved with ASP and OMP.

\begin{table*}[!t]
\centering
\small
\setlength{\tabcolsep}{8pt}
\caption{Energy and force mean absolute errors (MAE) obtained using different sparse regression packages on the silicon and molybdenum datasets~\cite{Zuo2020}. The ASP and OMP solvers are from {\tt ActiveSetPursuit.jl}~\cite{ActiveSetPursuitJL}, the LARS solver is from {\tt LARS.jl}~\cite{KornblithJohnson2025LARSjl}, and the LASSO path solver is from {\tt Lasso.jl}~\cite{JuliaStatsLasso}. All reported models were trained using an initial basis of 5292 functions and selected approximately 1000 active features.}
\label{tab:software_Zuo}

\begin{subtable}{\textwidth}
\centering
\caption{Mean absolute test errors in predicted energies (meV).}
\label{tab:software_Zuo_energies}
\begin{tabular}{@{}lcccccccc@{}}
\toprule
\textbf{Element} & \multicolumn{4}{c}{\textbf{Small model [meV]}} & \multicolumn{4}{c}{\textbf{Large model [meV]}} \\ 
\cmidrule(lr){2-5} \cmidrule(lr){6-9}
& ASP & OMP & LARS & LASSO & ASP & OMP & LARS & LASSO \\
\midrule
Mo & 2.261 & 2.123 & 20.627 & 21.066 & 2.138 & 1.933 & 12.892 & 13.787 \\
Si & 2.444 & 2.296 & 19.916 & 20.111 & 1.981 & 1.859 & 12.970 & 13.326 \\
\bottomrule
\end{tabular}
\end{subtable}

\vspace{0.6cm}

\begin{subtable}{\textwidth}
\centering
\caption{Mean absolute test errors in predicted forces (eV/\AA).}
\label{tab:software_Zuo_forces}
\begin{tabular}{@{}lcccccccc@{}}
\toprule
\textbf{Element} & \multicolumn{4}{c}{\textbf{Small model [eV/\AA]}} & \multicolumn{4}{c}{\textbf{Large model [eV/\AA]}} \\ 
\cmidrule(lr){2-5} \cmidrule(lr){6-9}
& ASP & OMP & LARS & LASSO & ASP & OMP & LARS & LASSO \\
\midrule
Mo & 0.104 & 0.098 & 0.247 & 0.254 & 0.087 & 0.086 & 0.152 & 0.159 \\
Si & 0.074 & 0.072 & 0.284 & 0.290 & 0.061 & 0.059 & 0.168 & 0.177 \\
\bottomrule
\end{tabular}
\end{subtable}
\end{table*}

\subsection{Silicon (PRX, 2018)}\label{sec:silicon18}
We employed the ASP and OMP solvers to fit a linear ACE potential to the silicon dataset presented in~\cite{PhysRevX.8.041048}, originally developed for training a Gaussian approximation potential (GAP). This dataset is more diverse and distinct from the Si dataset used in Section~\ref{sec:limit_metal}. This dataset is comprehensive, encompassing a diverse range of configurations, including various bulk crystal structures (e.g., diamond, hcp, and fcc), amorphous phases, and liquid molecular dynamics (MD) snapshots. Its goal is to capture a wide range of the elemental silicon energy landscape. Accurate fits to this dataset provide (near-)general-purpose potentials for elemental silicon.
For this benchmark, the data are randomly divided into 70\% training, 15\% validation, and 15\% test sets. The split respects training structure categories, such as different phases or defects.

Four models were evaluated: the GAP model and three ACE models trained with default parameters. The ACE model trained using BLR employed basis functions with a maximum correlation order of $N_{\text{max}} = 3$, a total polynomial degree of $D_{\text{tot}} = 20$, and a cutoff radius of $r_c = 6\, \text{\AA}$, resulting in a basis size of 5,456. 

Additionally, two more ACE models were trained using ASP and OMP, both with the same maximum correlation order of 3 but a higher total degree of \( D_{\text{tot}} = 23 \), yielding a larger initial basis size of 13,695. In both cases, training was stopped once 5,000 active basis functions had been selected. An algebraic smoothness prior $\Gamma$ with $p = 5$ was employed in all ACE models.
A key property of a robust interatomic potential is its out-of-distribution generalization performance. To evaluate this, the benchmark suite of property predictions from~\cite{PhysRevX.8.041048} was performed. It includes bulk diamond elastic constants, vacancy formation energies, surface formation energies for the (100), (110), and (111) surfaces, as well as formation energies for hexagonal, dumbbell, and tetragonal point defects in bulk diamond. The reference values for these properties were obtained via CASTEP~\cite{ClarkSegallPickardHasnipProbertRefsonPayne} density functional theory (DFT) calculations. 

For comparison, we included results from the GAP potential trained in~\cite{PhysRevX.8.041048} and the ACE BLR potential trained in~\cite{10.1063/5.0158783}. The computed values of these properties are summarized in Table~\ref{tab:Sicomparison}. 

\begin{table*}[!t]
\centering
\scriptsize  
\footnotesize
\setlength{\tabcolsep}{3.0pt}
\renewcommand{\arraystretch}{0.95}
\caption{Comparison of point defects, elastic properties, and surface energies for silicon using DFT, GAP, BLR, ASP, and OMP models (cf.~\ref{sec:silicon18}). Models were fitted to the dataset in~\cite{PhysRevX.8.041048}, with reference values obtained from CASTEP~\cite{ClarkSegallPickardHasnipProbertRefsonPayne} density functional theory (DFT) calculations.}
\label{tab:Sicomparison}
\begin{tabular}{c|cccc|cccc|ccc}
\hline
\textbf{Method} &
\multicolumn{4}{c}{\textbf{Point Defects [eV]}} &
\multicolumn{4}{c}{\textbf{Elastic Properties [GPa]}} &
\multicolumn{3}{c}{\textbf{Surface Energy [J/m$^2$]}} \\
\cline{2-12}
& Tet. Int. & Hex. Int. & Dumb. Int. & Vac. & $B$ & $c_{11}$ & $c_{12}$ & $c_{44}$ & (111) & (110) & (100) \\
\hline
DFT         & 3.91 & 3.72 & 3.66 & 3.67 & 88.6 & 153.3 & 56.3 & 72.2  & 1.57 & 1.52 & 2.17 \\
GAP         & 3.71 & 3.63 & 3.68 & \textbf{3.53} & \textbf{88.5} & \textbf{151.9} & 63.1 & \textbf{99.1} & \textbf{1.54} & \textbf{1.55} & \textbf{2.13} \\
BLR         & 3.74 & 3.61 & 3.72 & 3.52 & 88.1 & 156.4 & 60.9 & 102.1 & 1.52 & 1.88 & \textbf{2.13} \\
ASP (2500)  & 3.70 & 3.66 & 3.73 & 3.51 & 88.3 & 157.7 & \textbf{59.9} & 101.3 & 1.52 & 1.87 & \textbf{2.13} \\
OMP (5000)  & \textbf{3.72} & 3.63 & \textbf{3.62} & 3.51 & 88.2 & 155.9 & 60.2 & 102.1 & 1.52 & 1.88 & \textbf{2.13} \\
ASP (5000)  & \textbf{3.72} & \textbf{3.67} & 3.78 & \textbf{3.53} & 88.3 & 156.6 & 60.2 & 102.0 & 1.52 & 1.87 & \textbf{2.13} \\
\end{tabular}
\end{table*}

These results demonstrate that the ASP and OMP solvers achieve accuracy comparable to the GAP potential and the ACE model trained using BLR, even when training is stopped at 2,500 basis functions—less than $50\%$ of the 5,456 basis functions used by the BLR-trained model.

\begin{figure*}[htb]
\centering
\begin{subfigure}{\textwidth}
  \centering  \includegraphics[width=0.9\linewidth]{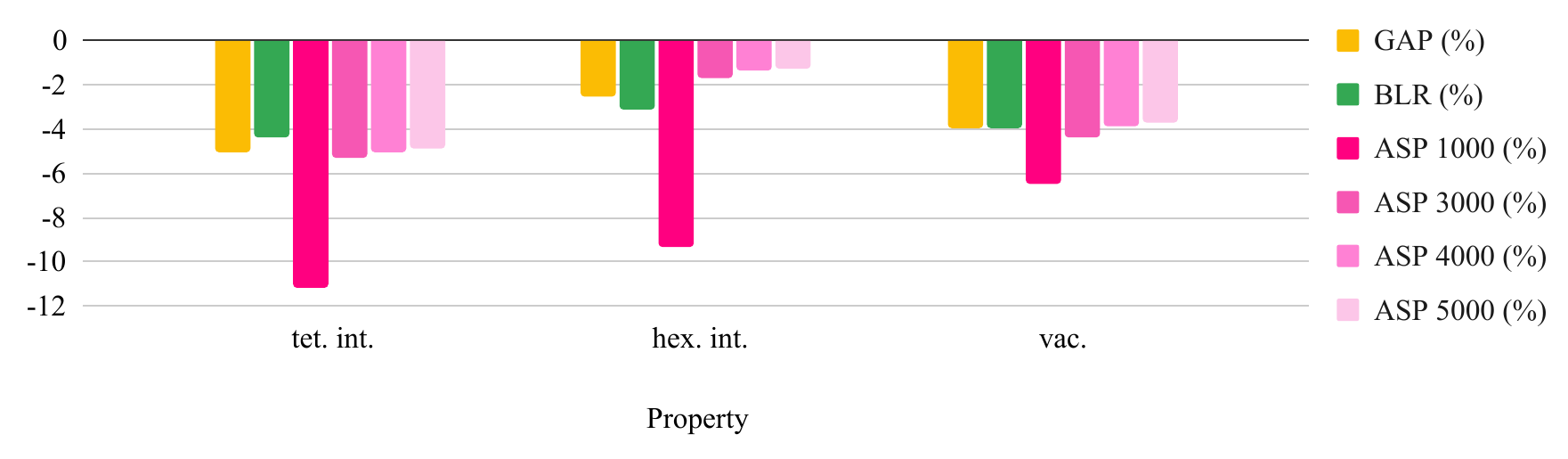} 
  \label{fig:sub1}
\end{subfigure}%
\caption{Percentage error relative to the computed values in Table~\ref{tab:Sicomparison} for various silicon properties using GAP, BLR, and ASP (cf. \ref{sec:silicon18}). The plot illustrates how the relative error of ASP decreases as the number of active basis functions increases, demonstrating a clear trend of improved accuracy.}
\label{fig:Sichart_lr}
\end{figure*}

The corresponding percentage errors of point defect energies relative to the DFT reference values are presented in Figure~\ref{fig:Sichart_lr}. The test illustrates how the accuracy of the model improves as the number of basis functions increases. This monotonicity of the error highlights the robustness of  {ASP} for basis selection. Again we observe comparable or better accuracy than {BLR} while using significantly fewer basis functions. The result further underscores the advantage of sparsity-promoting solvers in balancing computational cost and predictive accuracy, demonstrating that a well-chosen subset of basis functions is sufficient for reliable property predictions.

\begin{figure}[H]
\centering
\includegraphics[width=0.4\textwidth]{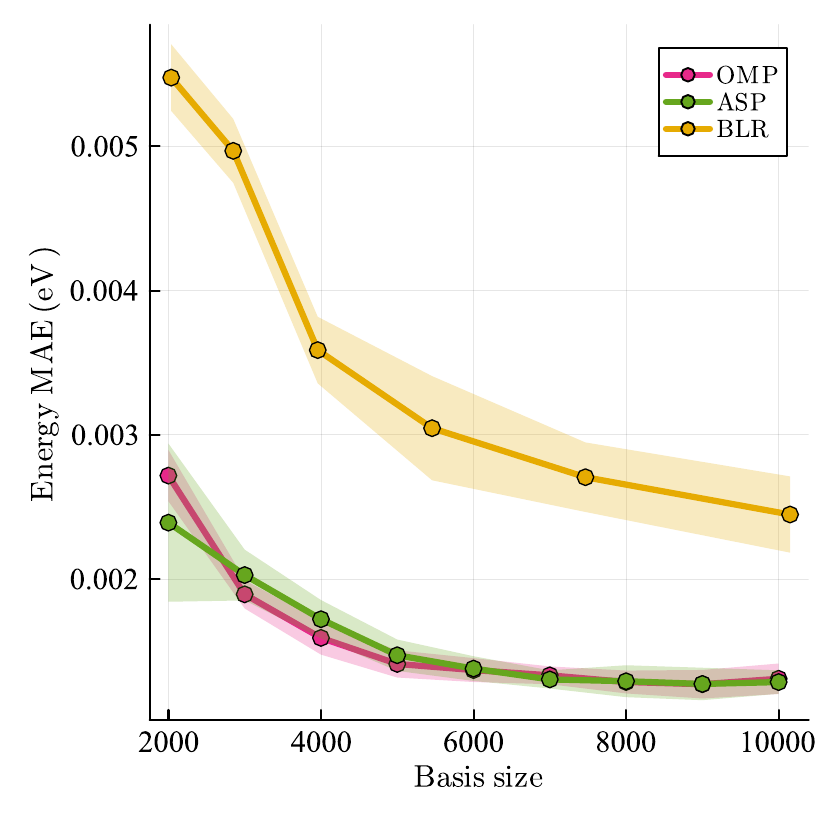}
\caption{Energy MAE as a function of basis size for the silicon dataset~\cite{PhysRevX.8.041048}. Mean test errors are averaged over 7 train–test splits, with shaded regions indicating 95\% confidence intervals for the mean. ASP and OMP outperform BLR while using significantly fewer basis functions.}
    \label{fig:Si-lrr}
\end{figure}

Figure~\ref{fig:Si-lrr} compares OMP, ASP, and BLR fits for silicon. For OMP and ASP, the solvers are initialized with a basis of 18,353 functions, and the test error is evaluated along the solution path up to 10,000 active basis functions. Results are averaged over 7 independent train–test splits, and shaded regions indicate 95\% confidence intervals for the mean test error.
In the small-basis regime, BLR and ASP exhibit comparable performance in terms of test error. As the basis size increases, the  {ASP} and  {OMP} solver consistently achieve moderately lower test errors compared to  {BLR}, which indicates that selecting an effective subset of basis functions leads to better generalization and less risk of overfitting, even in the current setting with a highly diverse dataset.


The basis selection patterns produced by the BLR and ASP solvers for the silicon dataset are similar to those observed in the previous section. Detailed visualizations can be found in the \textit{Supplementary Information}.

\subsection{Water}\label{sec:water}
We utilize OMP and ASP to fit an interatomic potential to a dataset containing 1593 liquid water configurations~\cite{doi:10.1073/pnas.1815117116}. 
In this case, we can test the selection of how basis functions acting between different elements are selected.
The model employed default parameters, including basis functions up to a correlation order of $N_{\text{max}} = 3$, a maximum polynomial degree of $D_{\text{tot}} = 15$, and a cutoff radius of $r_{\text{c}} = 5.5$ Å. An algebraic smoothness prior $\Gamma$ with $p = 4$ was employed in all ACE models.

\begin{figure}[htb!]
\centering
\includegraphics[width=0.4\textwidth]{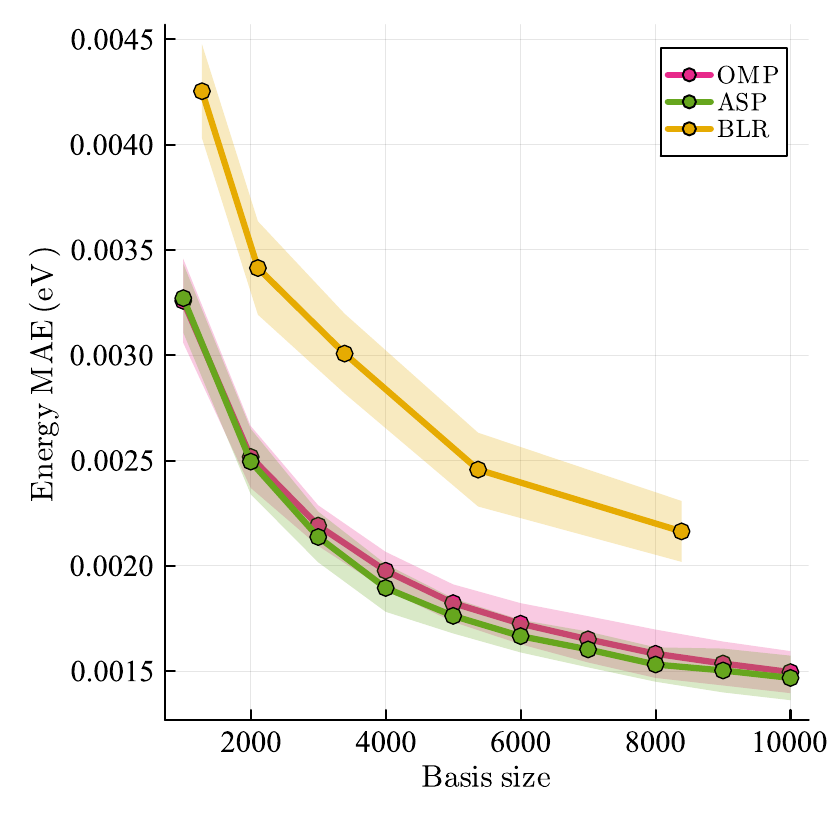}
\caption{Energy MAE vs. basis size for the water dataset~\cite{doi:10.1073/pnas.1815117116}, comparing OMP, BLR, and ASP (cf.~\ref{sec:water}). ASP and OMP outperform BLR while using significantly fewer basis functions. Solid lines indicate the mean error over 7 train–test splits, with shaded regions indicating 95\% confidence intervals for the mean.}
    \label{fig:water-LR}
\end{figure}

We compared the performance of the ASP, OMP, and BLR solvers using 7 independent random splits of the dataset, each partitioned into 70\% training, 15\% validation, and 15\% test sets. Figure~\ref{fig:water-LR} shows the learning curves for ACE models trained on the water dataset, starting from an initial basis of 19,501 functions, with the test error evaluated along the solution path up to 10,000 active basis functions. Results are averaged over the 7 splits, and shaded regions indicate 95\% confidence intervals for the mean. OMP and ASP achieve comparable MAE and consistently outperform BLR. As OMP offers similar predictive accuracy while being more computationally efficient, we present all further results using only the OMP solver.


We next compared the OMP solver with Cartesian Atomic Cluster Expansion (CACE)~\cite{cace2024}, which is a recently proposed framework that formulates atomic density expansions directly in Cartesian coordinates, bypassing the need for spherical harmonics. 

In the following numerical test, the OMP model was initialized with a basis of 19,501 functions, and training was terminated once the number of active basis functions reached 3,000, 9,000, and 12,000.

For a fair comparison, the dataset~\cite{doi:10.1073/pnas.1815117116} was partitioned into 1,354 training configurations and 239 test configurations, which were used to train and evaluate the OMP, BLR, and CACE models. A validation set comprising 15\% of the training data was used for model selection.

Table~\ref{tab:water_errors} compares the results obtained using OMP and  {BLR} with CACE. CACE \(T=0\) relies on local atomic features (similar to linear ACE), while \(T=1\) incorporates message passing to capture non-local interactions, improving accuracy at a higher computational cost.

Notably, the ACE models fitted using OMP require significantly fewer parameters than other models in Table~\ref{tab:water_errors}. Specifically, the OMP solver achieves lower energy error than the CACE ($T=0$) model while using less than 50\% of the parameters. This is despite being a linear model, whereas CACE employs nonlinear tensor decompositions and a learnable radial basis. The larger force error suggests that tuning the regression weights may be beneficial, a process that should arguably also be automated in future work. 

\begin{table*}[htb!]
    \centering
    \small
    \setlength{\tabcolsep}{6pt}
    \renewcommand{\arraystretch}{1.2} 
    \caption{Test Mean Absolute Errors (MAE) for energy (E) per water molecule (in meV/H$_2$O) and force (in meV/\AA) for models trained on a liquid water dataset from Ref.~\cite{doi:10.1073/pnas.1815117116}. The number of message passing layers \(T\) for the CACE models are listed. All models used a cutoff radius of $r_\text{cut} = 5.5 \text{\AA}$. Despite being a linear model, the sparse solver achieves comparable accuracy with significantly fewer parameters, demonstrating the effectiveness of data-driven basis selection.
    }
    \label{tab:water_errors}
    
    \begin{tabular}{l c c c c} 
        \hline
        \textbf{Model}  & \textbf{Energy MAE} (meV/atom) & \textbf{Force MAE} (meV/\AA) & \textbf{\# Parameters} \\ 
        \hline
        CACE~\cite{cace2024} (\(T=1\))    & 0.89  & 27  & 69320  \\ 
        CACE~\cite{cace2024} (\(T=0\))   & 1.42  & 42  & 24572  \\ 
        ACE ({BLR})~\cite{10.1063/5.0158783}  & 1.86  & 76  & 19501  \\
 
                ACE ({OMP})  
        & 1.58 & 71 & 6000 \\
        ACE ({OMP}) & 1.44  & 65  & 9000  \\
        ACE ({OMP}) & 1.36  & 62  & 12000  \\
        \hline
    \end{tabular}
\end{table*}

To test the stability of trained potentials, we ran NVT MD simulations at 300 K with a timestep of 1 fs for a total simulation time of 300 ps. From these simulations, we computed the diffusivities reported in Table~\ref{tab:Diffusivities}.

\begin{table}[htb]
    \centering
    \small
    \setlength{\tabcolsep}{6pt}
    \renewcommand{\arraystretch}{1.2} 
\caption{Diffusivities obtained from NVT MD simulations at 300~K using different potentials, compared with the DFT reference value. Overall, diffusivity errors follow the same trend as force errors, with CACE ($T=0$) being the only exception.}
    \label{tab:Diffusivities}
    
    \begin{tabular}{c c c c c} 
        \hline
        \textbf{Model} & \textbf{Diffusivity ({\AA}$^2$/ps)} & \textbf{Relative Diffusivity error} \\
        \hline
                ACE BLR & 0.120 & 0.47 \\ 
                ACE OMP (6000) & 0.101 & 0.55\\ 
        ACE OMP (9000) & 0.118 & 0.47 \\ 
        ACE OMP (12000) & 0.129 & 0.42\\ 
        CACE ($T=0$)  & 0.168 & 0.25\\
        CACE ($T=1$)  & 0.174  & 0.23\\
DFT~\cite{Marsalek2017-gl} & 0.226  & -\\ 
        \hline
    \end{tabular}
\end{table}

As the number of active basis functions increases, the computed diffusivity approaches the DFT value. The diffusivity errors follow essentially the same trend as the force errors, with models that achieve lower force MAEs reported in Table~\ref{tab:water_errors}, generally producing more accurate predictions. 
The diffusivity accuracy of linear ACE models is generally poor compared with the nonlinear CACE models, which is consistent with previous studies~\cite{10.1063/5.0158783}. 
Most importantly for our purposes is that all potentials are sufficiently MD stable to be able to predict diffusivity and that the OMP potential matches or outperforms the dense BLR ACE potential in accuracy with significantly fewer active basis functions.

\begin{figure*}[htb!]
     \centering
     
     \begin{subfigure}[b]{0.3\linewidth}
         \centering
         \includegraphics[width=\textwidth]{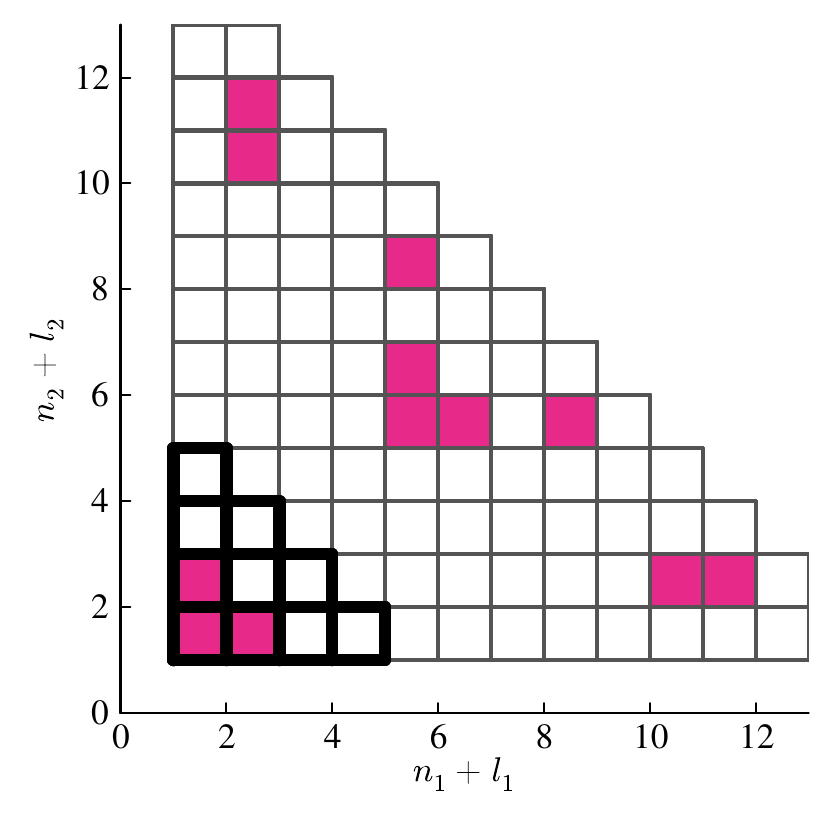}
         \caption{O-O-H: \#active = 100}
         \label{fig:wat2d_HH_100}
     \end{subfigure}
     \hfill
     \begin{subfigure}[b]{0.3\linewidth}
         \centering
         \includegraphics[width=\textwidth]{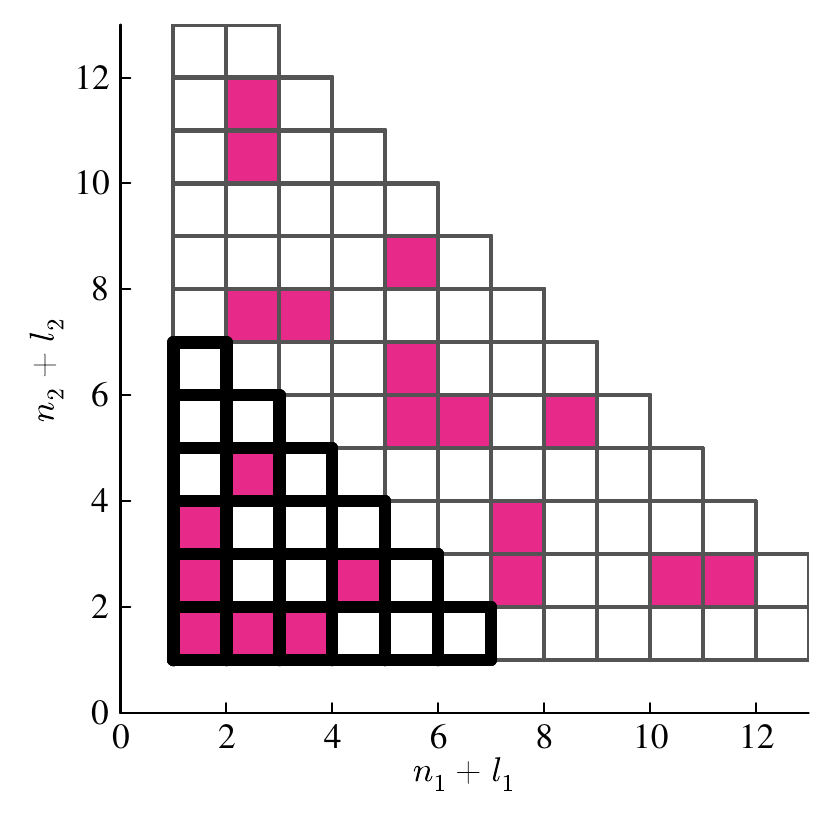}
         \caption{O-O-H: \#active = 500}
         \label{fig:wat2d_HH_500}
     \end{subfigure}
     \hfill
     \begin{subfigure}[b]{0.3\linewidth}
         \centering
         \includegraphics[width=\textwidth]{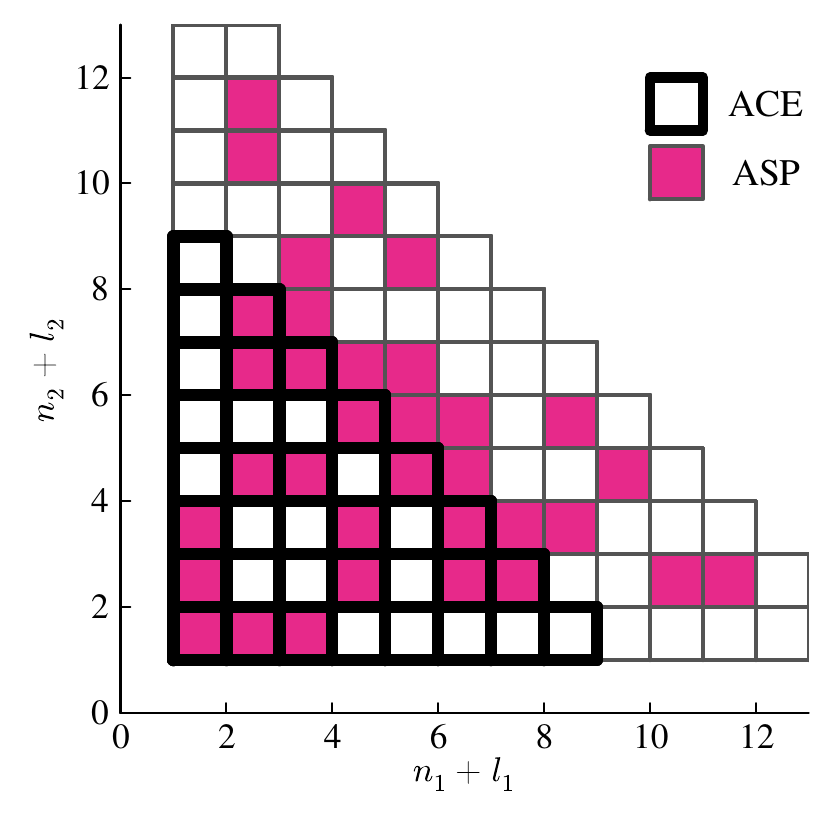}
         \caption{O-O-H: \#active = 1200}
         \label{fig:wat2d_HH_1200}
     \end{subfigure}
     \vspace{0.1cm}
     
     \begin{subfigure}[b]{0.3\linewidth}
         \centering
         \includegraphics[width=\textwidth]{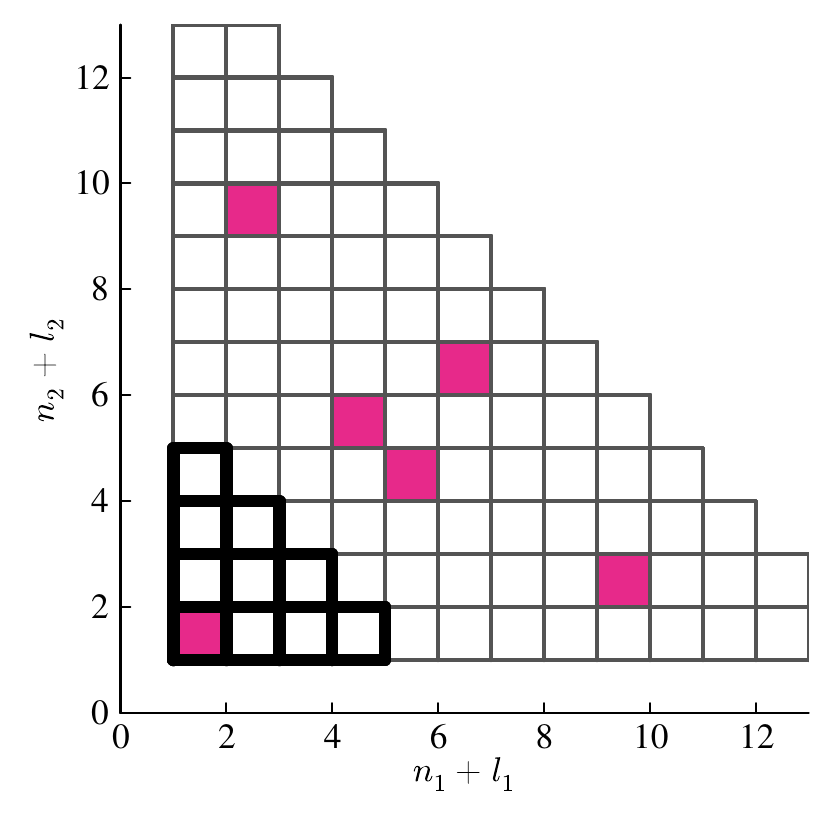}
         \caption{O-O-O: \#active = 100}
         \label{fig:wat2d_OO_100}
     \end{subfigure}
     \hfill
     \begin{subfigure}[b]{0.3\linewidth}
         \centering
         \includegraphics[width=\textwidth]{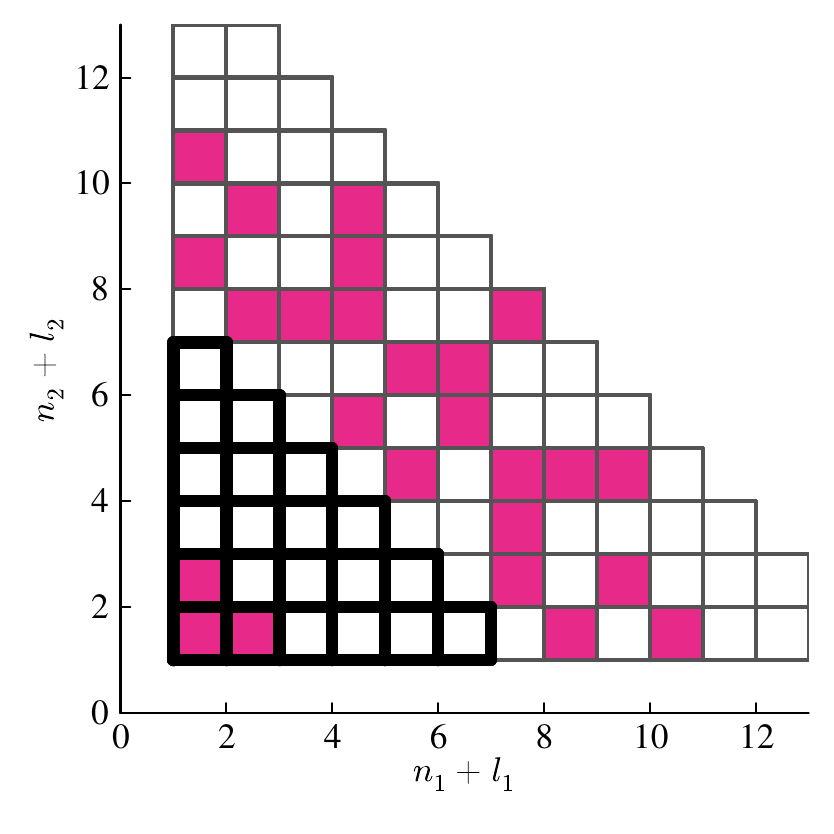}
         \caption{O-O-O: \#active = 500}
         \label{fig:wat2d_OO_500}
     \end{subfigure}
     \hfill
     \begin{subfigure}[b]{0.3\linewidth}
         \centering
         \includegraphics[width=\textwidth]{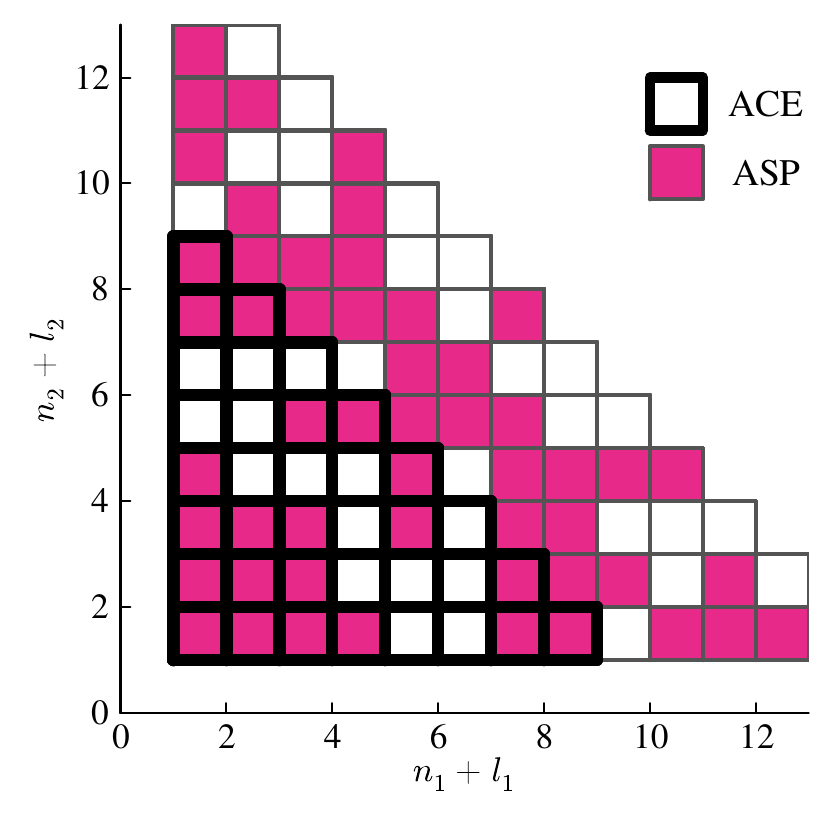}
         \caption{O-O-O: \#active = 1200}
         \label{fig:wat2d_OO_1200}
     \end{subfigure}
     \vspace{0.1cm}
     
     \begin{subfigure}[b]{0.3\linewidth}
         \centering
         \includegraphics[width=\textwidth]{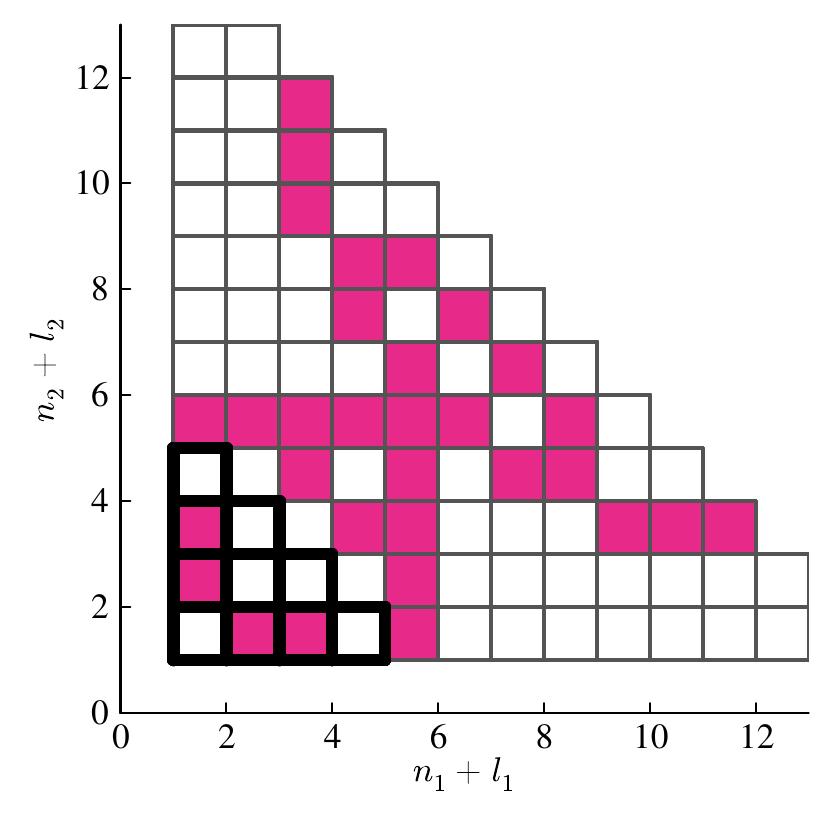}
         \caption{O-H-H: \#active = 100}
         \label{fig:wat2d_HO_100}
     \end{subfigure}
     \hfill
     \begin{subfigure}[b]{0.3\linewidth}
         \centering
         \includegraphics[width=\textwidth]{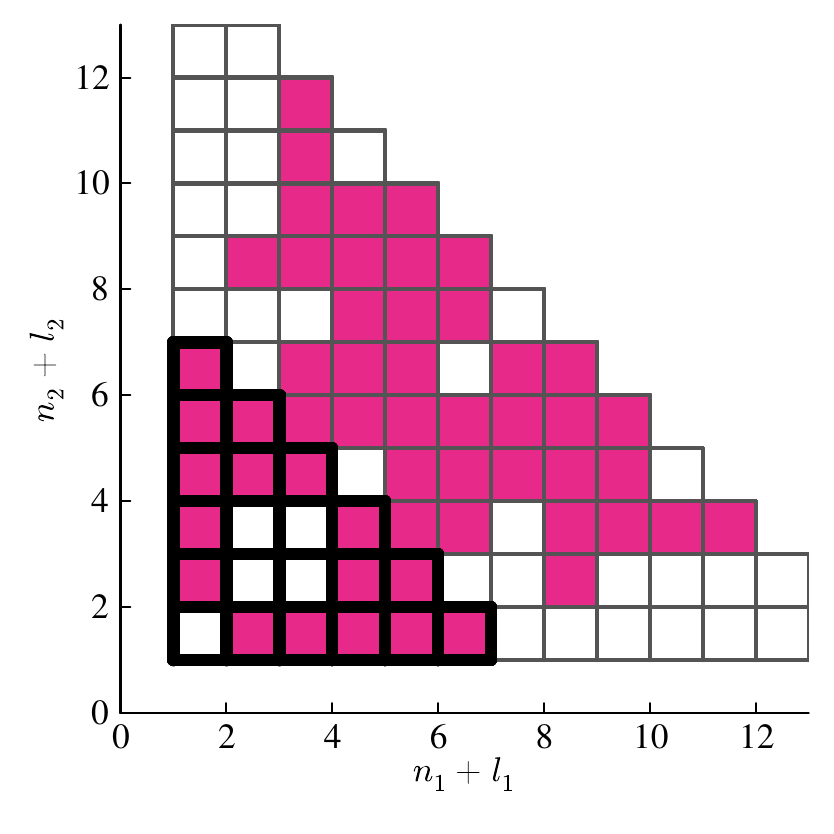}
         \caption{O-H-H: \#active = 500}
         \label{fig:wat2d_HO_500}
     \end{subfigure}
     \hfill
     \begin{subfigure}[b]{0.3\linewidth}
         \centering
         \includegraphics[width=\textwidth]{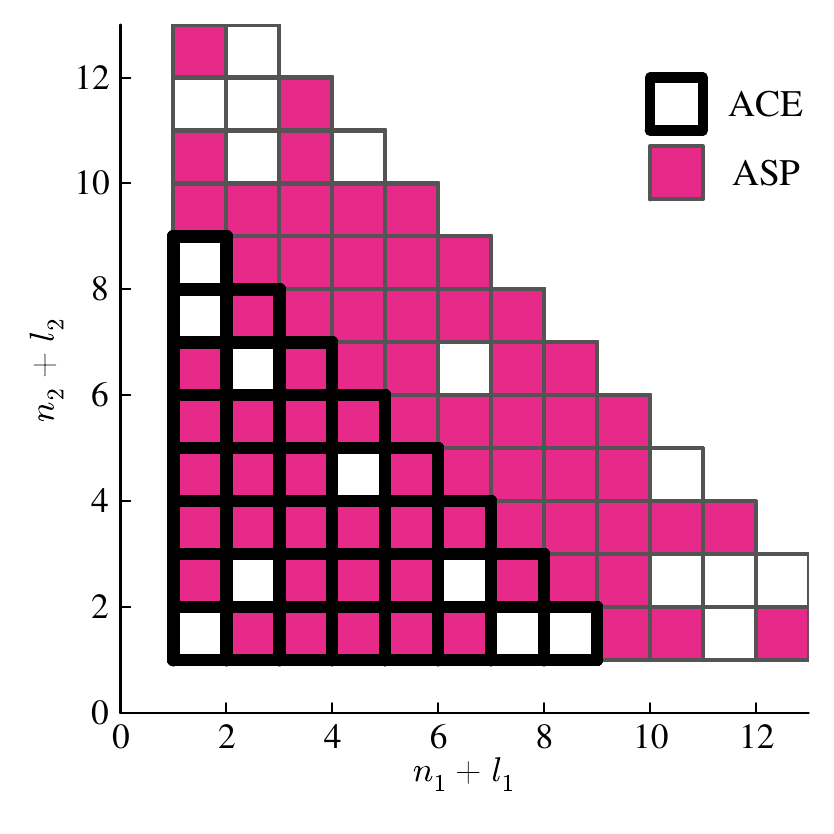}
         \caption{O-H-H: \#active = 1200}
         \label{fig:wat2d_HO_1200}
     \end{subfigure}
     \vspace{0.5cm}
     \caption{Illustration of ASP and other (non-sparse) ACE solvers' gradual selection of three-body basis functions for the water dataset~\cite{doi:10.1073/pnas.1815117116}, with O as the center atom (cf.~\ref{sec:water}). Each row corresponds to a different interaction type: O-H-H, O-O-O, and O-O-H. The results highlight the greater importance of resolving O-H interactions for accurately modeling the potential energy surface. This aligns with chemical intuition, as these bonds dominate water’s structural and energetic properties due to their role in hydrogen bonding and intermolecular interactions.}
     \label{fig:water_combined}
\end{figure*}

To illustrate the behavior of sparse solvers in this setting, Figure~\ref{fig:water_combined} shows the basis functions selected by the ASP solver for three-body interactions, with oxygen (O) as the central atom.

The selection pattern prioritizes O-H-H interactions over O-O-O and O-O-H, which intuitively should indeed have a the most important contribution to the total energy and forces. This selection pattern demonstrates that sparse, data-driven basis selection successfully identifies and prioritizes the most physically relevant interactions. Rather than following a hierarchical scheme, the solver autonomously captures the underlying physics of water, emphasizing the O-H-H bonds which are the building blocks of the water molecule~\cite{Titantah2013}. Since we already have prior intuition about the interactions in water, the selected pattern mainly serves to confirm it. However, it also illustrates the potential of sparse solvers to identify the most relevant interactions in systems where such intuition is lacking, helping to identify the most important interactions directly from the data.

\subsection{Effects of correlation order}
\begin{figure}[H]
\centering
     \begin{subfigure}[b]{0.4\linewidth}
         \centering
        \includegraphics[width=\textwidth]{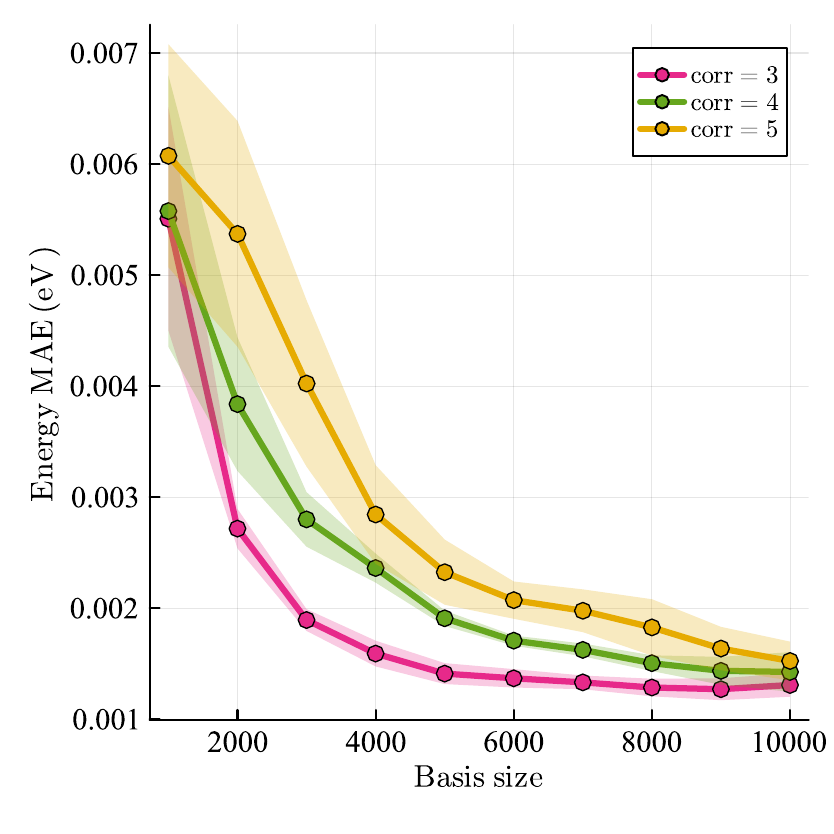}
         \caption{}
         \label{fig:si-std-comp}
     \end{subfigure}
    \vspace{0.4cm}
         \begin{subfigure}[b]{0.4\linewidth}
         \centering
\includegraphics[width=\textwidth]{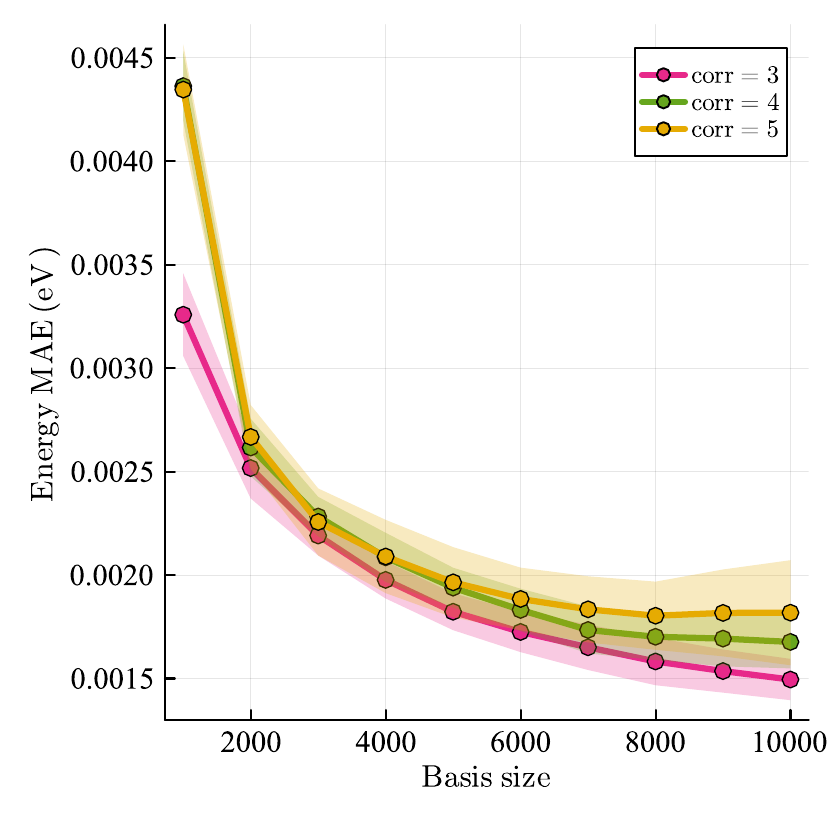}
         \caption{}
         \label{fig:water-std-comp}
     \end{subfigure}
    \caption{Energy MAE as a function of basis size for OMP models trained with different correlation orders \(N_{\max}=3,4,5\). (a) shows results for the Silicon dataset~\cite{PhysRevX.8.041048}, and (b) for the water dataset~\cite{doi:10.1073/pnas.1815117116}. In both systems, increasing the correlation order does not yield a consistent improvement in test-set accuracy at fixed sparsity. Solid lines indicate the mean error over 7 train–test splits, with shaded regions indicating 95\% confidence intervals for the mean.}

    \label{fig:corr-orders}
\end{figure}

To examine the role of the physics encoded in the descriptor set, we performed additional sparse OMP fits using correlation orders $N_{\max}=3,4,5$ for both silicon and water. For each case, the polynomial degree was chosen such that the total number of candidate basis functions was comparable (approximately \(2\times10^{4} \pm 2\times10^{3}\)). For each correlation order, we trained 7 models using random train-test splits; with shaded
regions indicating 95\% confidence intervals for the mean. The resulting learning curves, shown in Fig.~\ref{fig:corr-orders}, demonstrate that our method is robust as the underlying dimensionality of the model increases. Comparable accuracy is achieved across correlation orders once an appropriate sparse subset of features is selected.

On the other hand, increasing the correlation order does not lead to a reduction in the error at fixed sparsity for either system. This behavior is in fact unsurprising: As shown in Figs.~\ref{fig:mo2d} and~\ref{fig:water_combined}, the sparse solvers tend to select high-degree basis functions at the lower correlation orders. When the correlation order is increased while keeping the total basis size fixed, these high-degree basis functions are no longer available in the candidate pool. As a result, increasing the correlation order does not improve accuracy at fixed sparsity.

\section{Conclusion}\label{sec:conclusion}

We explored the use of sparse machine learning interatomic potentials (MLIPs) using active-set algorithms for automated data-driven feature selection. By integrating the \texttt{ActiveSetPursuit.jl} package into the Atomic Cluster Expansion (ACE) framework, we constructed sparse linear models that achieve both improved accuracy relative to dense ACE models as well as substantially reducing model complexity. The computational tests confirm that sparse optimization substantially reduces the need for manual hyperparameter tuning while improving model generalization. 

For the datasets used in this study, OMP and ASP demonstrate similar accuracy-cost trade-offs on average. OMP, being a greedy algorithm that only adds elements to the active set, is somewhat more efficient in terms of wall-clock time. ASP, on the other hand, offers a more principled path-tracing approach, as it follows the exact $\ell_1$-regularized solution path rather than relying on greedy updates. Both outperform the other sparse optimization methods we benchmarked, achieving better results in terms of both cost and accuracy. Looking forward, OMP has the potential advantage that it can be more easily extended to nonlinear regression settings.

It's noteworthy that our approach is not restricted to single-component systems. Because it builds on sparse regression applied to a chosen feature set, the framework can be readily transferred to other materials once a suitable basis is defined. This makes the method broadly applicable, from simple elemental systems to complex alloys and compounds. The key limitation is that for systems with {\em many} chemical species (say, more than 4) the added combinatorial complexity makes linear model inherently expensive. In this case one should apply our approach to a linear model in which species are embedded in a low to moderate-dimensional latent space. The challenge to overcome is that such embeddings are typically ``learned'', which would result in a nonlinear parameter estimation.

\section{Acknowledgements}
The work was supported by the New Frontiers in Research Fund through an NFRF Exploration Grant, and by the Natural Sciences and Engineering Research Council of Canada through an NSERC Discovery grant and an NSERC-NSF Collaboration on quantum science and artificial intelligence grant. CO is a partner in Symmetric Group LLP which licenses force fields commercially. The authors acknowledge helpful discussions with Gábor Csányi and Cas van der Oord. This research was supported in part through computational resources and services provided by Advanced Research Computing at the University of British Columbia.

\section{Appendix}

\subsection{Radial Basis Functions}\label{app:ACE}
The general form of this one-particle basis is given by:
\[
\phi_{\bm n \bm \ell \bm m}(x_{ij}) = P_{\bm n \bm l}(r_{ij}, Z_i, Z_j) Y_{\bm \ell}^{\bm m}(\hat{\bm{r}}_{ij}),
\]
where  \(Y_{\bm \ell}^{\bm m}\) denote the complex or real spherical harmonics, while $P_{\bm n \bm l}$ are radial basis functions with significant freedom of choice.
The angular component $\hat{\bm{r}}_{ij}$ is embedded in the spherical harmonics, ensuring the parameterization is symmetric with respect to rotations. 
The radial basis function $R_{\bm n \bm l}(r_{ij}, Z_i, Z_j)$ is essential in modeling atomic interactions, and in the  \texttt{ACEpotentials.jl} package, these bases are indexed only by $\bm n$, meaning $R_{\bm n \bm l} = R_{\bm n}$ for all $\bm l$. This function is defined as

\begin{equation}
R_{\bm n}(r_{ij}, Z_i, Z_j) = f_{\text{env}}(r_{ij}, Z_i, Z_j) P_{\bm n}(y(r_{ij}, Z_i, Z_j)), \nonumber
\end{equation}

where \( y \) is an element-specific distance transformation designed to enhance spatial resolution, particularly near equilibrium bond lengths. A common form is
\begin{equation}
y(r_{ij}, Z_i, Z_j) = \left( 1 + a \left( \frac{r}{r_0} \right)^q \right) \left( 1 + \left( \frac{r}{r_0} \right)^{q - p} \right)^{-1},\nonumber
\end{equation}

with \( r_0 \) as an estimate of the equilibrium bond length and \( a \) chosen to maximize the gradient of \( y \) at \( r = r_0 \). Here, \( P_{\bm n} \) is an orthogonal polynomial basis in the \( y \)-coordinates, typically chosen as Legendre polynomials to ensure even resolution over the \( y \)-domain. The function \( f_\text{env} \) serves as an envelope, specifying a cutoff radius \( r_\text{cut} \). For the many-body basis, a typical choice is
\begin{equation}
f_\text{env}(r_{ij}, Z_i, Z_j) = y^2 (y - y_\text{cut})^2, \nonumber
\end{equation}
where \( y_\text{cut} = y(r_\text{cut}, Z_i, Z_j) \).

For the pair potential $V^{(1)}$, the default envelope is a modified Coulomb potential which ensures repulsive behavior as $r_{ij} \to 0$, while being continuously differentiable at the cutoff.
\begin{align}
    f_\text{env}(r_{ij}, Z_i, Z_j) &= \left( \frac{r_{ij}}{r_0} \right)^{-1} - \left( \frac{r_\text{cut}}{r_0} \right)^{-1} \nonumber + \\& \left( \frac{r_\text{cut}}{r_0} \right)^{-2} \left( \frac{r_{ij}}{r_0} - \frac{r_\text{cut}}{r_0} \right). \nonumber
\end{align}

For more information on the Atomic Cluster Expansion framework, the reader should refer to~\cite{10.1063/5.0158783}.

\subsection{The BPDual Algorithm}\label{sec:bpdual}

The BPDual algorithm proposed by Friedlander and Saunders~\cite{friedlander2012dual}, is an active-set method specifically tailored for solving a class of quadratic programs (QP) that arise as the dual of Basis Pursuit Denoising (BPDN) problem. In this section we will briefly review the main algorithm proposed. For more details the reader should refer to~\cite{friedlander2012dual}.

The optimal solutions of the basis pursuit denoising (BPDN)~\ref{eq:primal-dual} and its dual QP are represented by $\mathbf{x}^\ast$ and $\mathbf{y}^\ast$, respectively. he objective function of the dual QP and its gradient are given by: 
\begin{equation}
\Phi(\mathbf{y}) = \frac{1}{2} \lambda ||\mathbf{y}||_2^2 - \mathbf{b}^T \mathbf{y}, \quad \mathbf{g}(\mathbf{y}) = \lambda \mathbf{y} - \mathbf{b} \nonumber
\end{equation}
If $\lambda > 0$, $\Phi(\mathbf{y})$ is strictly convex, ensuring a unique solution $\mathbf{y}^\ast$. The vectors $\mathbf{a}_j$ and $\mathbf{e}_j$ represent the j-th columns of the matrix $\mathbf{A}$ and the identity matrix, respectively. The j-th component of a vector $\mathbf{z}$ is denoted by $z_j$. The support of a vector, denoted by $\mathcal{S}$, is the set of indices $j$ where the corresponding element is non-zero: $x_j \neq 0$. The subvector $\mathbf{x}_\mathcal{S}$ contains only the non-zero components of the vector.

The BPDual algorithm is an active-set method. The active set method focuses on identifying constraints that are active (i.e., binding) at each iteration. For an estimated solution \( \textbf{y} \), a constraint is active if \( \textbf{a}_j^\top \textbf{y} \) lies at one of its bounds (i.e. $\textbf{a}_j^\top \textbf{y} = \mathbf{1}$ or $-\mathbf{1}$). The active set \( S \) is a set of indices representing the active constraints at each iteration, which forms a submatrix of \( \textbf{A} \) that has full column rank. For instance, for some permuation $\mathrm{P}$
\begin{equation}
\mathrm{\textbf{A}P = \left[ S \; N \right] \quad \text{and} \quad S^\top \textbf{y} = \pm \mathbf{1}}.\nonumber
\end{equation}

The estimate \( \textbf{y} \) is stationary if it satisfies the constraints \( -\mathbf{1}\leq \textbf{A}^\top \textbf{y} \leq \mathbf{1} \) and the objective gradient \( \mathbf{g}(\mathbf{y})\) is a linear combination of active constraint gradients. Specifically, there exists \( x_S \) such that:
\begin{equation}
    S \mathbf{x}_\mathcal{S} + \lambda \textbf{y} = \textbf{b} \quad \text{and} \quad \mathrm{S^T \textbf{y} = \pm \mathbf{1}}. \label{eq:stationary}
\end{equation}
These conditions imply that the current estimates \( (\textbf{x}, \textbf{y}) \) satisfy feasibility for the primal problem, with \( \textbf{x} \) residing within the active constraints.

At each iteration, the algorithm seeks directions \( \Delta \mathbf{x}_\mathcal{S} \) and \( \Delta \mathbf{y} \) such that the updated \( \mathbf{x}_\mathcal{S} \) and \( \mathbf{y} \) are stationary (i.e., they are feasible and satisfy the stationarity conditions~\ref{eq:stationary}). This is achieved by solving:
\begin{equation}
\min_{\Delta \mathbf{x}_\mathcal{S}} \| \mathbf{h} - S \Delta \mathbf{x}_\mathcal{S} \|_2, \quad \Delta \mathbf{y} = \frac{\mathbf{h} - S \Delta \mathbf{x}_\mathcal{S}}{\lambda},\nonumber
\end{equation}

where \( \mathbf{h} = \mathbf{b} - \lambda \mathbf{y} - S \mathbf{x}_\mathcal{S} \).

\begin{figure}[H]
\centering
\begin{subfigure}{.5\textwidth}
  \centering
\includegraphics[width=\linewidth]{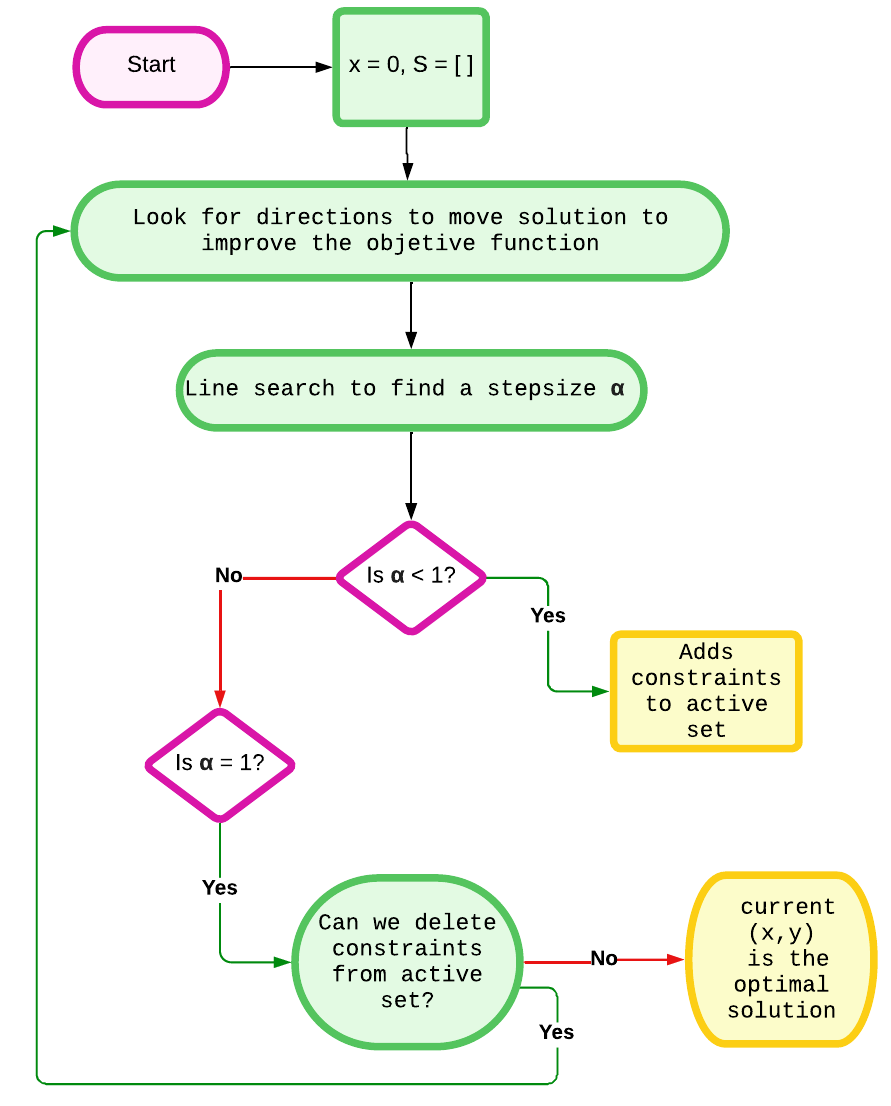}
\end{subfigure}
\caption{A schematic of the BPDual algorithm.
}
\label{fig:asp-chart}
\end{figure}

An overview of the BPDual algorithm is provided in Figure~\ref{fig:asp-chart}.
It's not worthy that for efficient computations, only the upper triangular part of a QR factorization of \( S \) is updated at each iteration using the \texttt{QRupdate.jl}~\cite{QRupdate.jl} package.

In the next step, a line search is performed to choose a step length \( \alpha > 0 \) to update \( \mathbf{x}_\mathcal{S} \) and \( \mathbf{y} \) by taking steps \( \mathbf{x}_\mathcal{S} \rightarrow \mathbf{x}_\mathcal{S} + \alpha \Delta \mathbf{x}_\mathcal{S} \) and \( \mathbf{y} \rightarrow \mathbf{y} + \alpha \Delta \mathbf{y} \), aiming to minimize \( \phi(\mathbf{y}) \). The step length \( \alpha = 1 \) is selected if possible, otherwise a constrained step is taken to avoid bound violations:
\begin{equation}
\alpha = \min \left\{ 1, \min_{j \notin S} \frac{\text{bound - } z_j}{\Delta z_j} \right\}.\nonumber
\end{equation}

If \( \alpha < 1 \), the line search encounters a constraint not currently in \( S \). This constraint is added to \( S \), ensuring that each new constraint introduced is linearly independent of the others in \( S \). The linearly independence criterion is enforced as any newly encountered constraint must have a normal vector not lying in the span of \( S \).

When \( \alpha = 1 \), the solution \( \textbf{y} + \Delta \textbf{y} \) minimizes \( \phi(\textbf{y}) \) on the current active set \( S \). If improvement is possible by removing a constraint from \( S \), the algorithm considers directions \( \Delta \textbf{y} \) that move \( \textbf{y} \) away from an active constraint, maintaining feasibility. 
Consider an inequality constraint \( q = S_p \) that is currently active at its lower bound, i.e., \( a_q^T \textbf{y} = - \mathbf{1} \). To move away from this bound, we select a direction \( \Delta \textbf{y} \) satisfying \( S^T \Delta \textbf{y} = e_p \), ensuring \( \Delta \textbf{y} \) is a feasible direction that moves away from the active lower bound. Observe that:
\begin{equation}
g^T \Delta \textbf{y} = (-S \mathbf{x}_\mathcal{S})^T \Delta \textbf{y} = -\mathbf{x}_\mathcal{S}^T S^T \Delta \textbf{y} = -(\mathbf{x}_\mathcal{S})_p.\nonumber
\end{equation}

Thus, if \( (\mathbf{x}_\mathcal{S})_p > 0 \), moving in the direction \( \Delta \textbf{y} \) will improve the objective, allowing us to remove the constraint \( q \) from the active set.

Similarly, if the constraint \( q \) is active at its upper bound and \( (\mathbf{x}_\mathcal{S})_p < 0 \), we can improve the objective by moving in a direction \( \Delta \textbf{y}\) for which \( S^T \Delta \textbf{y} = -e_p \). In this case, we also remove constraint \( q \) from the active set. Hence, the Lagrange multipliers \( \mathbf{x}_\mathcal{S} \) indicate which indices in the active set \( S \) should be removed to enable a reduction in the objective function. If none of the elements in \( \mathbf{x}_\mathcal{S} \) have the appropriate sign, then the current point \( (\textbf{x}, \textbf{y}) \) is optimal.

\subsection{Lasso Path}\label{sec:homotopy}
The homotopy approach, is a special case of parametric programming~\cite{linoptimBertsimas}, aims to solve a sequence of LASSO problems, gradually adjusting a regularization parameter to trace a continuous path of solutions.

To apply the homotopy method to the dual problem~\ref{eq:primal-dual} begins with a large initial value of \( \lambda = \|\textbf{A}^\top b\|\), where the solution \( \mathbf{x} \) is fully sparse. The parameter \( \tau \) is then gradually decreased, following a continuous "path" of solutions that progressively allow more non-zero entries in \( \mathbf{x} \). This path of solutions is often referred to as the \emph{Lasso coefficient path}. 
At each iteration \( k \), the algorithm solves a Lasso problem with the updated parameter \( \lambda_k \) via the BPDual algorithm. 

As \( \lambda_k \) decreases, more coefficients in \( \mathbf{x} \) move from zero to non-zero values, revealing additional active variables in the model. 
Suppose \( \mathbf{y} \) is the optimal solution to the dual problem for a given \( \tau \), and let \( \mathbf{x}_S \) represent the vector of Lagrange multipliers for the active set \( S \). The objective is to find the maximum reduction in \( \lambda \) such that \( S \) remains the optimal active set. Letting \( \bar{\lambda} = \lambda - \alpha \), the corresponding optimal pair \( (\bar{\mathbf{x}}_S, \bar{\mathbf{y}}) \) satisfies the equations:

\begin{equation}
S \bar{\mathbf{x}}_S + \lambda \bar{\mathbf{y}} = \mathbf{b} \quad \text{and} \quad S^T \bar{\mathbf{y}} = \pm \textbf{1}. \label{eq:homotopyopt}
\end{equation}

The homotopy algorithm calculates the largest allowable decrease in \( \lambda \) while maintaining feasibility of the current active set. 
To determine the direction in which the solution will change, we subtract the original optimality condition~\ref{eq:stationary} from~\ref{eq:homotopyopt} and use the definition \( \bar{\lambda} = \lambda - \alpha \), yielding:

\begin{equation}
    S(\bar{\mathbf{x}}_S - \mathbf{x}_S) + \lambda(\bar{\mathbf{y}} - \mathbf{y}) - \alpha \mathbf{y} = 0, \quad S^T (\bar{\mathbf{y}} - \mathbf{y}) = 0.\nonumber
\end{equation}

By pre-multiplying the first equation by \( S^T \) and applying the second equation, we obtain:

\begin{equation}
    S^T S (\bar{\mathbf{x}}_S - \mathbf{x}_S) = \alpha S^T \mathbf{y}.\nonumber
\end{equation}

This equation can be interpreted as the optimality condition for the following least-squares problem:

\begin{equation}
    \bar{\mathbf{x}}_S = \mathbf{x}_S + \alpha \Delta \mathbf{x}, \quad \text{where } \Delta \mathbf{x} = \underset{\Delta \mathbf{x}}{\text{argmin}} \|\mathbf{y} - S \Delta \mathbf{x}\|_2.\nonumber
\end{equation}

The direction of change for \( \mathbf{x}_S \) is thus independent of  \( \alpha \). The maximum step \( \alpha_j^x \) that can be taken without changing the sign of any component \( \mathbf{x}_j \) is given by:

\begin{equation}
\alpha_j^\mathbf{x} = \begin{cases} 
    -\mathbf{x}_j / \Delta \mathbf{x}_j & \text{if } j \in \mathcal{C}, \\
    +\infty & \text{otherwise},\nonumber
\end{cases} 
\end{equation}

where \(
\mathcal{C} = \{ j \mid \operatorname{sign}(\mathbf{x}_j) = -\operatorname{sign}(\Delta \mathbf{x}_j) \). Thus, the feasible step size \( \alpha \) cannot be larger than \( \alpha^\mathbf{x} = \min_j \alpha_j^\mathbf{x} \).

As \( \lambda \) decreases to \( \bar{\lambda} \), it is also necessary to assess the impact on \( \mathbf{y} \), particularly with respect to the constraints \( -\mathbf{1} \leq A^T \mathbf{y} \leq \mathbf{1}\). Subtracting \( \bar{\mathbf{y}} \) from both sides and using the definition of \( \Delta \mathbf{x} \), we obtain:

\begin{equation}
(\lambda - \alpha)(\bar{\mathbf{y}} - \mathbf{y}) = \mathbf{b} - \lambda \mathbf{y} - S \mathbf{x}_S + \alpha (\mathbf{y} - S \Delta \mathbf{x}).\nonumber
\end{equation}

Due to the optimality of the current pair \( (\mathbf{x}, \mathbf{y}) \), we have:

\begin{equation}
\bar{\mathbf{y}} = \mathbf{y} + \frac{\alpha}{\lambda - \alpha} \Delta \mathbf{y}, \quad \text{where } \Delta \mathbf{y} := \mathbf{y} - S \Delta \mathbf{x}.\nonumber
\end{equation}

This expression represents the residual in the least-squares problem. Consequently, the change in constraints can be captured by:

\begin{equation}
\bar{\mathbf{z}} = \mathbf{z} + \frac{\alpha}{\lambda - \alpha} \Delta \mathbf{z}, \quad \text{where } \mathbf{z} := A^T \mathbf{y} \quad \text{and} \quad \Delta \mathbf{z} := A^T \Delta \mathbf{y}.\nonumber
\end{equation}

The maximum feasible step \( \alpha_j^\mathbf{z}  \) without violating a constraint \( j \notin S \) is:

\begin{equation}
\alpha_j^\mathbf{z}  = 
\begin{cases}
    \frac{\lambda (1 - \mathbf{z}_j)}{\Delta \mathbf{z}_j - \mathbf{z}_j + 1} & \text{if } \Delta \mathbf{z}_j > 0, \\
    \frac{\lambda (-1 - \mathbf{z}_j)}{\Delta \mathbf{z}_j - \mathbf{z}_j -1} & \text{if } \Delta \mathbf{z}_j < 0, \\
    +\infty & \text{otherwise}, \nonumber
\end{cases} 
\end{equation}

Thus, the feasible step size \( \alpha \) cannot exceed \( \alpha^\mathbf{z} = \min_{j \notin S} \alpha_j^\mathbf{z} \). The total allowable reduction in \( \lambda \) is therefore \( \alpha = \min \{\alpha^\mathbf{x}, \alpha^\mathbf{z}\}\).

In each iteration, the homotopy algorithm updates the active set and adjusts the solution \( (\mathbf{x}, \mathbf{y}) \) by taking the largest feasible step \( \alpha \), driving the optimization towards minimizing the objective function while adhering to feasibility constraints.

\subsection{Orthogonal Matching Pursuit}\label{sec:omp}
Orthogonal Matching Pursuit (OMP) is a greedy algorithm for finding a sparse solution to the linear system \( Ax = b \). OMP constructs the support of \( x \) iteratively, adding one nonzero component at each step. Initially, the solution vector is set to \( x_0 = 0 \), the active set is \( \mathcal{S}_0 = \emptyset \), and the residual is \( r_0 = b \). At each iteration \( k \), OMP identifies the column \( a_{i} \) of \( A \) that is most correlated with the current residual \( r_{k-1} \). Specifically, it selects the index \( i \) that maximizes the absolute value of the following inner product, \( i = \underset{i}{\arg \max} | \langle a_i, r_{k-1} \rangle | \). The active set is then updated as \( \mathcal{S}_k = \mathcal{S}_{k-1} \cup \{ i_k \} \). The algorithm computes the new solution \( x_k \) by minimizing the residual norm \( \| b - A x \|_2 \) over vectors supported on \( S_k \). Finally, the residual is updated as \( r_k = b - A x_k \). The algorithm enforces the orthogonality of \( r_k \) with respect to the columns of \( A \) in the current support, ensuring that once a column is selected, it will not be chosen again. This process repeats until a stopping criterion, such as a sufficiently small residual norm, is met.

\subsection{BRR Posterior Derivation}\label{app:brr}
Starting with the Gaussian likelihood and prior, the posterior distribution for the coefficients $x$ is derived using Bayes' theorem:
\[
  p(x \mid b, A, \alpha, \lambda) \propto p(b \mid A, x, \lambda) \, p(x \mid \alpha).
\]
Detailed algebra shows that the posterior is Gaussian with covariance
\[
  \boldsymbol{\Sigma} = (\alpha \mathbf{I} + \lambda A^\top A)^{-1},
\]
and mean
\[
  \hat{x} = \lambda \boldsymbol{\Sigma} A^\top b.
\]

The marginal likelihood (evidence) is given by:
\begin{equation}
  p(b \mid A, \alpha, \lambda) = \int p(b \mid x, A, \lambda) \, p(x \mid \alpha) \, dx.\nonumber
\end{equation}
Taking the logarithm yields:

\begin{align}
  \ln p(b \mid A, \alpha, \lambda) =& -\frac{m}{2}\ln(2\pi) + \frac{1}{2}\ln |\boldsymbol{\Sigma}| \nonumber \\ 
  &- \frac{\lambda}{2}\|b - A\hat{x}\|^2 - \frac{\alpha}{2}\|\hat{x}\|^2. \nonumber
\end{align}

This marginal log-likelihood is maximized with respect to $\alpha$ (or $\boldsymbol{\alpha}$ in the ARD case) and $\lambda$, typically using optimization methods such as BFGS or L-BFGS.

\subsection{Truncated Singular Value Decomposition (TSVD) Post-Processing} \label{app:tsvd}

The  {ASP} solver applies $\ell_1$-regularization, which promotes sparsity but can also shrink the recovered coefficients excessively toward zero. To mitigate this effect and improve numerical stability, we apply a post-processing step using truncated singular value decomposition (TSVD). 

Given the selected subset of basis functions \( S \), let \( A_S \) denote the reduced design matrix corresponding to these basis functions, and let \( x \) be the final iterate obtained from  {ASP}. The TSVD-based refinement is performed as follows:
\begin{enumerate}
    \item Compute the singular value decomposition (SVD) of the matrix \( A_S \):
    \begin{equation}
        A_S = U \Sigma V^T,\nonumber
    \end{equation}
    where \( U \in \mathbb{R}^{m \times r} \) and \( V \in \mathbb{R}^{n \times r} \) are orthonormal matrices, and \( \Sigma \in \mathbb{R}^{r \times r} \) is a diagonal matrix containing the singular values \( \sigma_1 \geq \sigma_2 \geq \dots \geq \sigma_r \).
    
    \item Define a truncation threshold \( \sigma_{\text{min}} \), typically chosen as a fraction of the largest singular value (e.g., \( \sigma_{\text{min}} = \epsilon \sigma_1 \) for some small \( \epsilon \)). Retain only the singular values satisfying \( \sigma_i > \sigma_{\text{min}} \), and form the truncated SVD:
    \begin{equation}
        A_S^{(\text{TSVD})} = U_k \Sigma_k V_k^T,\nonumber
    \end{equation}
    where \( k \) is the number of singular values exceeding \( \sigma_{\text{min}} \), and \( U_k, \Sigma_k, V_k \) are the corresponding truncated matrices.

    \item Project the solution \( x \) onto the lower-rank subspace:
    \begin{equation}
        x_{\text{TSVD}} = V_k \Sigma_k^{-1} U_k^T b.\nonumber
    \end{equation}
\end{enumerate}

By applying TSVD, we filter out small singular values that introduce numerical instability while preserving the dominant components of \( x \). This step counteracts the over-shrinkage induced by the \(\ell_1\)-regularization and ensures a more stable and accurate recovery of the coefficient vector.

\bibliographystyle{plain}
\bibliography{bib.bib}

\end{document}